\documentclass[lettersize,journal]{IEEEtran}
\IEEEoverridecommandlockouts
\usepackage{float}
\usepackage{comment}
\makeatletter
\let\newfloat\newfloat@ltx
\makeatother
\usepackage[utf8]{inputenc}
\usepackage{xcolor}
\usepackage{graphicx}
\usepackage{amsmath}
\usepackage{amsthm}
\usepackage[normalem]{ulem}
\usepackage{amssymb}
\usepackage[hyphens]{url}
\usepackage{soul}
\usepackage{subfig}
\newcommand\redout{\bgroup\markoverwith
{\textcolor{red}{\rule[.5ex]{2pt}{0.4pt}}}\ULon}
\usepackage[ruled,linesnumbered,vlined]{algorithm2e}
\usepackage{etoolbox}
\usepackage{nomencl}
\usepackage{authblk}
\usepackage{comment}
\usepackage{array}
\PassOptionsToPackage{hyphens}{url}
\usepackage{hyperref}
\usepackage{balance}

\newcounter{algoline}
\newcommand\Numberline{\refstepcounter{algoline}\nlset{\thealgoline}}
\AtBeginEnvironment{algorithm}{\setcounter{algoline}{0}}
\usepackage{booktabs}
\usepackage{multirow}
\let\originalleft\left
\let\originalright\right
\renewcommand{\left}{\mathopen{}\mathclose\bgroup\originalleft}
\renewcommand{\right}{\aftergroup\egroup\originalright}

\newcommand{\Prob}[1]{\mathrm{Pr}\left[#1\right]\xspace}
\newcommand{\Expt}[1]{\mathbb{E}\left[#1\right]\xspace}

\newcommand{\para}[1]{\vspace{0.1em}\noindent\textbf{#1.}~}

\def\BibTeX{{\rm B\kern-.05em{\sc i\kern-.025em b}\kern-.08em
    T\kern-.1667em\lower.7ex\hbox{E}\kern-.125emX}}
\begin{document}

\title{Exposing Hidden Attackers in Industrial Control Systems using Micro-distortions 
\thanks{A preliminary version of this work has appeared in \cite{9632311}.}
}

\author{\IEEEauthorblockN{Suman Sourav}}
\author{\IEEEauthorblockN{Binbin Chen}}

\affil{Singapore University of Technology and Design, Singapore; e-mail: \{suman\_sourav, binbin\_chen\}@sutd.edu.sg}

\maketitle

\begin{abstract}

For industrial control systems (ICS), many existing defense solutions focus on detecting attacks only when they make the system behave anomalously. Instead, in this work, we study how to detect attackers who are still in their hiding phase. Specifically, we consider {an off-path false-data-injection attacker who makes the original sensor's readings unavailable
and then impersonates that sensor by sending out legitimate-looking fake readings,} 
so that she can stay hidden in the system for a prolonged period of time (e.g., to gain more information or to launch the actual devastating attack on a specific time). To expose such hidden attackers, our approach relies on continuous injection of ``micro distortion'' to the original sensor's readings, either through digital or physical means. We keep the distortions strictly within a small magnitude (e.g., $0.5\%$ of the possible operating value range) to ensure that it does not affect the normal functioning of the ICS. Micro-distortions are generated based on secret key(s) shared only between the targeted sensor and the defender. 
For digitally-inserted micro-distortions, we propose and discuss the pros and cons of a two-layer least-significant-bit-based detection algorithm. Alternatively, when the micro-distortions are added physically, a main design challenge is to ensure the introduced micro-distortions do not get overwhelmed by the fluctuation of actual readings and can still provide accurate detection capability.
Towards that, we propose a simple yet effective {\it Filtered-$\Delta$-Mean-Difference} algorithm that can expose the hidden attackers in a highly accurate and fast manner. We demonstrate the effectiveness and versatility of our defense by using real-world sensor reading traces from different industrial control (including smart grid) systems. 
\end{abstract}

\section{Introduction}

Given the central role of Industrial Control Systems (ICS) in different critical infrastructures like smart grids \cite{mcdaniel2009security}%
, water %
treatment systems \cite{weiss2014industrial}%
, and nuclear power plants \cite{cho2015cyberphysical}, the issue of ICS security has become increasingly important. 
High-profile attacks like Stuxnet~\cite{nourian2015systems} and the Ukraine power grid blackouts~\cite{lee2017crashoverride} have shown that adversaries can wreak havoc by compromising ICS devices and manipulating their readings or behavior. In many cases, attackers who have already gained some foothold in the system choose to stay hidden for a prolonged period of time so as to %
launch the attack in a coordinated time (e.g., to maximize the attack's impact), {or to launch the ``frog boiling attack'' \cite{10.5555/3327345.3327509, DBLP:journals/corr/abs-2002-02741, 10.1145/1639562.1639592} without triggering detection mechanism.} Instead of reacting to the launch of an actual attack only when the attack causes the ICS to deviate from the expected system behavior, we aim to expose hidden attackers proactively before they do any damage. 
{Specifically, we consider an attacker who wants to inject malicious sensor data into an ICS, however, she cannot put herself directly on-path (i.e., to gain direct control of the original sender or to launch a man-in-the-middle %
attack via ARP spoofing). We call such an attacker an off-path attacker and her attack  consists of two logical steps: (1) to make the original sensor unavailable (e.g., by crashing the sensor's firmware \cite{codesys} or through other forms of denial-of-service attack against the sensor \cite{wood2002denial}), (2) to impersonate the sensor and inject crafted data \cite{regan2017survey}. 
One such end-to-end attack is demonstrated by researchers from IOActive in Black Hat USA 2017 conference against a nuclear plant that follows a similar attack sequence \cite{wired, santamarta2017go}. We provide further discussion and justification of our threat model in Section \ref{sec:threat_model}. }
The attacker wants to remain hidden, so if there are any intrusion detection mechanisms in the ICS, the attacker will use her best knowledge about both the system's operation behavior and the intrusion detection rules to carefully craft the fake sensor readings, making them look normal and indistinguishable from the real sensor's readings.  

{While there have been significant advances in securing ICS against attacks on sensor readings (e.g.,~\cite{6545301, %
7011170, 7039974}),  %
implementation of many of these %
solutions require major upgrading of the existing ICS, e.g., by introducing authentication schemes at both the sensors and their corresponding receivers which can significantly amplify the cost of such upgrading.}
Other works \cite{10.1109/TIFS.2015.2512522, 8004369,10.1145/3274694.3274748} are based on assumptions that may not hold for advanced and persistent attackers, e.g., assuming the attackers do not know about the system's operational behavior or cannot observe some unique features of the sensors before launching the attack.

\begin{table*}[]
\centering
\begin{tabular}{|m{0.95cm}|m{3.75cm}|m{3.7cm}|m{8.1cm}|}
\hline
 Distortion & Advantages over other schemes & Additional advantages & Potential drawbacks \\ \hline
Digital & \multirow{2}{*}{\begin{tabular}[m{5cm}]{@{}l@{}}1) Smaller attack surface:\\ Secret only shared between the \\ protected sensor and the defender.\\\\ 2) Easy to deploy:\\ No change/upgrade required for \\ legacy receivers (e.g., controllers).\end{tabular}} & 

\begin{tabular}[l]{@{}l@{}}Extremely fast and accurate\\ detection: error rate reduced\\ exponentially with the number\\ of readings.\end{tabular} & \begin{tabular}[l]{@{}l@{}}1) {The digital distorter needs to be connected to the %
network,}\\ {hence is an easier cyber attack target compared to physical distorter.}\\ 2) The secret bits of one-time pad are sent out explicitly over the air,\\ making it more vulnerable to weakness in the one-time pad used (this\\ risk can be mitigated by our two-layer design).\\ 3) More costly to introduce the digital distorter.
\end{tabular} \\ \cline{1-1} \cline{3-4} 
Physical &  & \begin{tabular}[l]{@{}l@{}}1) Physical distorter can be \\ better isolated from attackers.\\ 2) No transmission of secret bits.\end{tabular} & \begin{tabular}[l]{@{}l@{}}1) Slower detection than digital distortion approach (significant 100x\\ speed-up can be achieved using our filter-$\Delta$-mean-difference approach).
\\
2) Physical distortion may be hard to introduce in some settings.\end{tabular} \\ \hline
\end{tabular}
\caption{\small The comparison of digital and physical distortions, as well as other schemes.}
\vspace{-0.5cm}
\label{tab:dig_vs_phy}
\end{table*}

In this work, we seek to design a practical solution for fast and accurate attacker detection in legacy ICS. The solution should be based on assumptions that even advanced attackers cannot easily bypass, %
while making minimal changes to a legacy ICS. Furthermore, 
the solution should have a minimal impact on the functioning of all legacy devices in the ICS.
{\noindent \bf Injecting micro-distortion based on a sequence of secret:}
The cornerstone of our defense strategy is to 
continuously introduce a very small distortion (which we will call ``micro-distortion'' hereafter) to the readings of the sensor that we want to protect. 
This micro-distortion needs to be within a very low magnitude, such that the normal functioning of the ICS remains unaffected.  In other words, these distortions should be tolerable by the devices that rely on the sensor reading, and the system should behave almost identically both with or without the distortions.

To use the presence of such distortion to authenticate the sensor, we generate the distortion based on a secret that is shared only between the sensor and a defender. The secret contains a sequence of binary values 0 and 1 (i.e., a one-time pad), one value to be used for each time instance.

There are two potential ways to introduce micro-distortion to the reading of a sensor in an ICS: 
\begin{itemize}

\item {\bf Digitally:} To upgrade the sensor's firmware or add a bump-in-the-wire device between the sensor and the ICS network, so as to introduce the micro-distortion into the original sensor reading. While this requires some change to the sensors, it doesn't affect the other parts of ICS (e.g., controllers), which requires %
more overhead to upgrade.

    \item {\bf Physically:} To deploy a micro-actuator that will physically add the micro changes to the underlying system so that the original sensor will pick that up in its reading. %

\end{itemize}

In either case, the secret sequence shared between a sensor and the defender
forms the basis for the defender to distinguish between the real sensor and a fake one. Table \ref{tab:dig_vs_phy} compares the digital and physical addition of distortions and summarizes their advantages over existing solutions.

{\noindent \bf
Digital introduction of micro-distortion:}
By digital means, specific bits of the sensor readings can be directly manipulated to introduce the micro-distortion. Therefore, a natural choice is to set the least significant bit (LSB) based on the shared secret key to keep the distortions minimal. 
Such digital tweaking of the LSB is extremely effective for attacker detection (i.e., with exponentially reducing error rate). 
{However, because of the potential drawbacks described in Table \ref{tab:dig_vs_phy}, it is important to study physical addition of micro-distortion too.}

{\noindent \bf
Physical introduction of micro-distortion:}
In this case, one cannot directly set a bit of a sensor reading; instead, the distortion will be added to the natural variation of sensor readings.
Hence, one key challenge for effective detection 
is that, by design, the magnitude of the micro-distortion is much lower than the sensor's actual readings. As a result, it can be easily overwhelmed by the latter. 
For example, if the actual readings are drawn uniformly randomly and independently from all the possible range of values, and when the micro-distortion $\epsilon$ equals $0.5\%$ of the possible range, it 
will require more than 80,000 samples in order to reduce both the false positive (FP) and false negative (FN) rate below $1\%$. 
If the sensor reading is sent every minute, such an approach requires an unacceptable detection delay of around $2$ months.
To overcome this, we leverage the observation that sensor readings in many ICS (and power grids in particular) often change gradually in a significant fraction of time (i.e., consecutive measurements have small difference).
Based on this observation, we devise an effective {\it Filtered-$\Delta$-Mean-Difference} algorithm that is based on statistical gauges calculated over the consecutive change of the sensor readings, instead of the raw sensor reading sequence directly. 
We demonstrate the effectiveness of our defense using real-world sensor reading traces from different types of ICS --- two smart grid systems,
one Secure Water Treatment (SWaT) testbed,
and a couple of synthetic datasets. 
For the smart grid systems, 
our experiments confirm that our detection algorithm under physical distortion can detect hidden attackers in a highly accurate (with false positive and false negative rate at $1\%$ or even lower) and fast (i.e., using less than 100 samples) manner, achieving more than $100$ times gain in terms of the detection delay compared to the baseline detection approach. 
For the SWaT system, a similar false positive and false negative rate at $1\%$ or lower can be achieved using only 50 samples.  
%
%

\para{Summary of enhancement from conference version and main contributions}
This paper is an extension based on a preliminary version of our work published in a conference~\cite{9632311}, where we only consider physical addition of micro-distortions to the sensor readings. The main addition in this paper include: (1) We propose new methods for digital addition of micro-distortion. (2) We analyze the pros and cons of different proposed techniques, especially between digital insertion and physical insertions of micro-distortion.
(3) We provide more thorough evaluations and analysis based on new datasets, including %
SWaT dataset and a couple of synthetic datasets. Those new results demonstrate the versatility and effectiveness of the proposed detection algorithms and provide theoretical insights regarding the limitations and insights behind the different detection algorithms.
In particular, we establish the requirement of a filtration step in the detection algorithm for ICS where there can be some large instantaneous changes. 

Overall, this paper is significantly enhanced from our conference version, providing a new and a more comprehensive proposal of different approaches, and more detailed analysis and experimental studies of the proposed approaches.

As summarized in Table \ref{tab:dig_vs_phy},
our micro-distortion-based solutions offer two promising candidates for an ICS defender to choose from. Whether digital or physical distortion is more suitable depends on the exact ICS setting, but overall they both offer a more effective, less costly, easier-to-deploy, and harder-to-bypass solutions compared to existing solutions, %
making them suitable for legacy ICS systems. 
In particular, the secret key needs to be shared only between the sensor and the defender.  Other components (e.g., controllers) can use the sensor's readings directly without requiring to filter out the injected distortions.
This not only reduces the chance of secret leakage but also makes deployability easier as none of these receiving components require upgrading.  %

%

%

%
%

%
%

%
\section{Related Work}

Attack detection in ICS, as opposed to traditional fault detection, is more challenging as the adversary here is usually persistent, intelligent, and stealthy. They can make use of the knowledge of the system to remain undetected \cite {7879966}. While attacks that manipulate the controller logic in ICS can be detected using software attestation~\cite{chen2017secure} and deception technology~\cite{mashima2017towards}, the existing solutions to counter sensor reading manipulations often face challenges when dealing with hidden attackers. For example, traditional bad-data detection techniques,  such as the largest residue test \cite{abur2004power},  may fail to detect an intelligent attacker who can change the state estimation of the system by introducing errors that fall within the range space of the observation matrix. Similarly, approaches such as \cite{biswas2019electricity} cannot work when redundant sensors that measure the same physical phenomenon are all compromised. %

Numerous detection techniques for false data injection attacks (FDIA) in power grids (e.g., \cite{10.1145/1952982.1952995, 7322210, 9069226, LI2019474}) rely on power grid state estimation and model the system as a stochastic linear/non-linear system that follows a strict mathematical modeling. In such cases, an attack is called stealthy if it is able to fool the system operator without being detected by a residue-based detector. %
To remain stealthy and avoid detection, the attacker needs to create false measurement data satisfying the constraints of state estimation.
If the attacker controls sufficient number of sensors to create self-consistent FDIA, or if it can mislead the control center state estimator to perceive a wrong operating state, 
the attacker would be able to bypass the existing detection mechanisms and cause large damage
\cite{7366616, 5622048}.
In comparison, the novelty of our proposed approach is that it is independent of the specific state of a system; rather it relies on micro-distortions based on a shared secret key for fast and accurate detection of hidden attackers.
Other works like \cite{9144530, en15197432, 9096314} etc. %
use computationally heavy and resource-consuming data-driven algorithms and machine learning (ML) models to detect false data. These techniques are orthogonal to our work in two ways. Firstly, our work relies on a shared secret key to create the micro-distortions, and a direct comparison against ML techniques that do not consider any shared secret key does not result in a fair comparison. Secondly, most of these techniques are computationally heavy and require costly resources and hardware, and hence these works go against our primary design goal of being low-cost.

Several watermarking-based authentication mechanisms where an actuator superimposes a random signal, known as the watermark, on the control policy-specified input while checking for an appropriate response from the sensors, were studied in \cite {5394956, 7011170}. %
There, the physical watermarking is added to the control output and studied specifically in the context of replay attacks, where an attacker just replays previously observed measurements of a system.  In \cite{7039974}, the watermarking scheme is extended for false data injection attacks where adversaries have the power to substitute real measurements with generated stealthy signals and was further improved in \cite{7738534,  8287297}.  %
Though similar in concept to a shared key, most of these solutions are focused on designing watermarked control inputs to detect counterfeit sensor outputs.  Moreover, these solutions are specific to linear dynamical systems described by time-invariant parameters and rely on accurate modeling of the system states.  
In contrast, our work pertains to detecting attackers by adding micro-distortions directly at the sensors which can then be leveraged to detect hidden attackers.
In \cite{10.1145/3274694.3274748, 10.1145/3196494.3196532}, the authors propose to authenticate sensors and detect data integrity attacks in CPSs by using a sensor's hardware characteristics along with physics of a process to create unique fingerprints for each sensor. Essentially, they create a noise fingerprint based on a set of time domain and frequency domain features that are extracted from the sensor and process noise. 
The main drawback of using such in-situ fingerprints is that such fingerprints can be learnt and potentially recreated by the attackers after the attacker makes sufficient observation. In comparison, in our work, we propose to artificially introduce micro-distortions based on a shared secret key between the distortion source and the detector (note that the original senders and receivers of those sensor readings do not need to be changed). The use of secret key removes the aforementioned drawback.
Furthermore, as shown in \cite{10.1145/3274694.3274748}, 
the false positive (FP) and false negative (FN) rate that approach can achieve is around $5\%$, which may not be acceptable for settings where any false positive or negative incurs a high cost to deal with. In comparison, even for a small observation window, the FP and FN rate of our approach is close to $0\%$.

Another relevant research area that we draw inspiration from is the line of work (see Chapter 6 of \cite{mohsenian-rad_2022} and references therein) that deals with the technique of \emph{probing}. In \cite{mohsenian-rad_2022}, probing is defined as the broad technique of perturbing a power system to enhance its monitoring capabilities; rather than just passively collecting measurements, perturbations are introduced into various grid components in order to actively create opportunities for gathering more knowledge about the power system. It is shown that introducing a perturbing (analog) signal can be useful for different tasks, including fault location identification in a power line, topology and phase identification of the underlying power grid, state and parameter estimation of power systems, etc. Similarly, we also use small perturbations (or micro-distortions in our term). However, our micro-distortion is based on a secret key, and the goal is to detect the presence of hidden attackers.

Different approaches based on cryptographic primitives have also been proposed to address this problem. For example,  homomorphic encryption based solutions were proposed in  \cite{kim2016encrypting,  7403296,  min2019privacy} and public-encryption systems in  \cite{6485982, 6857845}. %
They utilize computationally heavy encryption/decryption algorithms, which not only increase delay but also require high upgradation costs for legacy systems. Solutions like \cite{8340689,  SEDA} rely on the installation of additional specialized equipment which can increase the upgradation costs significantly.  
\addtolength{\topmargin}{0.05in}
\section{Threat Model} \label{sec:threat_model}

\begin{figure}[t!]
  \centering
    \includegraphics[width=\linewidth]{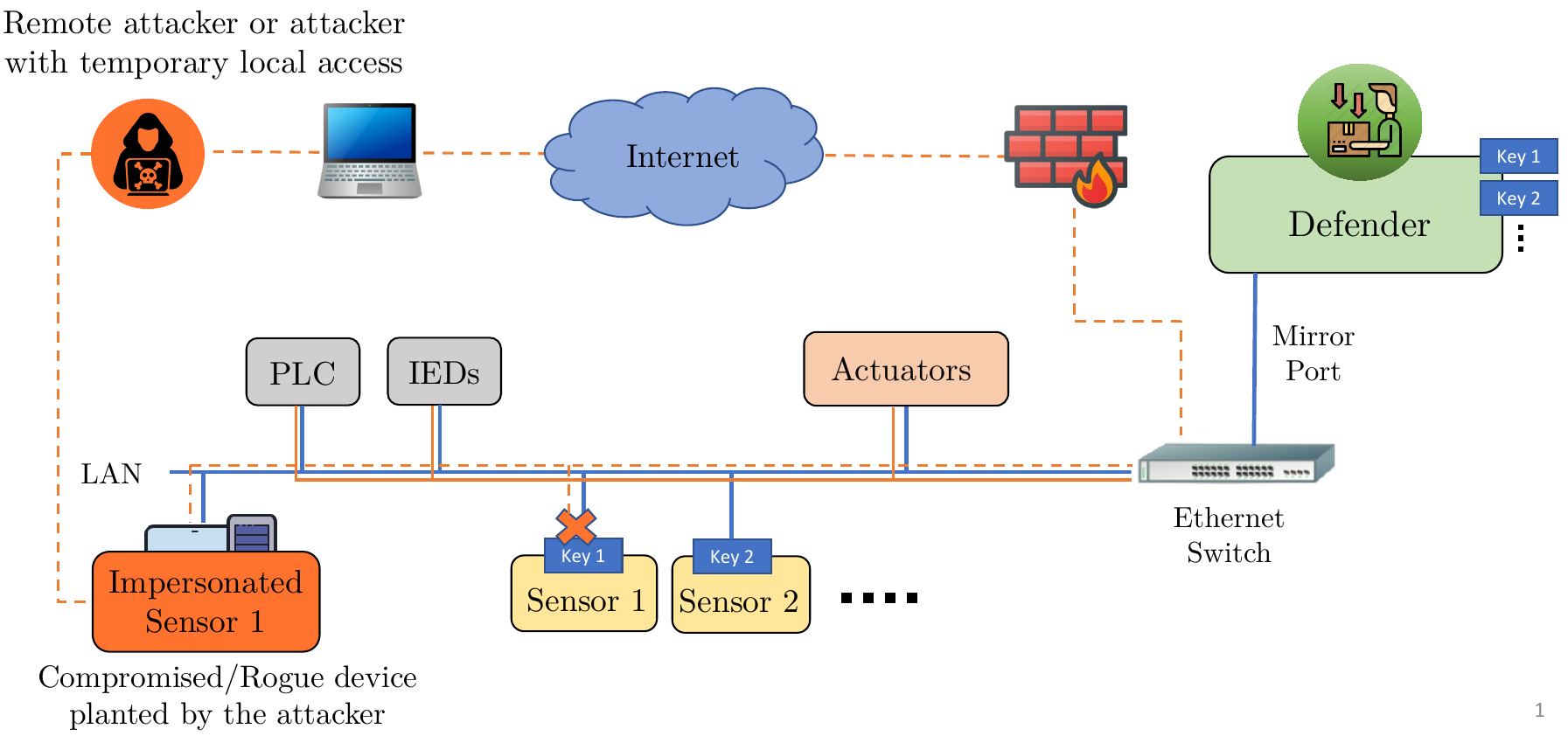}
  \caption{\small Our threat model over a typical ICS. The blue line represents the connection between the devices, the dashed red line represents the attack path where the remote attacker {(or attacker with temporary local access)} crashes Sensor 1 and she then impersonates the sensor (e.g., via a compromised device {or a rogue device brought in by the attacker}). The red line shows the traffic from the impersonated sensor to the other connected devices through the Ethernet switch. In the absence of an attack, similar traffic flow exists between Sensor 1 and the other devices.\vspace{-3mm}}
  \label{fig:system}
\end{figure}

As shown in Figure~\ref{fig:system}, 
%
%
%
%
%
%
%
{%
we consider a %
hidden attacker that
aims to (eventually) inject false sensor data to the ICS in a stealthy manner by first making the targeted sensor (or more specifically, the field device that sends out the sensor readings) unavailable and then impersonating the sensor by injecting crafted sensor data to the ICS.
} 

{An attacker can make a sensor unavailable in several different ways. Once the sensor is offline/crashes, any communication to and from the sensor is not possible. One common option is to crash a sensor by exploiting flaws in its software.} These flaws are usually more easily discoverable by attackers than flaws allowing an attacker to take complete control of a sensor, which is a strictly higher requirement for the attackers than just crashing the device. In fact, a search of programmable logic controller (PLC) related vulnerabilities in the Common Vulnerabilities and Exposures (CVE) database shows that in the year 2021, around 80\% of
newly reported vulnerabilities can cause Denial of Service/crashing of the PLC, while only around 20\% of vulnerabilities can potentially allow full control of the PLC. For example, see vulnerabilities like CVE-2021-22789 to CVE-2021-22792, etc. %
As a proof of concept, we exploited the recent CVE-2022-32137~\cite{codesys} vulnerability of the CODESYS V2 runtime and successfully crashed a PLC used in an energy research testbed. The CODESYS V2 runtime attacked here is widely used by many leading PLC manufacturers. We were not able to take full control of that PLC using the same vulnerability though. {There also have been several studies where the embedded devices can be crashed or made unavailable by purposely introducing faults through high voltage or electromagnetic fault injections \cite{quisquater2002eddy, 10.1145/3230833.3230842}, by explicit physical means (e.g., by simply disconnecting their power or network cable, or more dramatically in the case when a substation was attacked by gunmen) \cite{homeland, NPR, lat}, or by denial-of-service attack via networks (e.g., jamming the transmission \cite{wired,santamarta2017go} and the very recent Brokenwire attack against Combined Charging System \cite{kohler2022brokenwire}).} 

{Also, there have been several studies on impersonating a sensor to inject false sensor data (e.g., \cite{regan2017survey, 10.1007/11801412_8, 6468406}).
Such false data injection can be done in several ways, such as sending false data into exposed network ports (similar to Kaminsky's attack~\cite{kaminsky2009dns, liu2006dns}), through wireless signal injection~\cite{wired,santamarta2017go}, or application-layer injection through exposed web or database APIs, or through session hijacking \cite{5718158, meghanathan2014network}. }

{Note that our threat model does not deal with an on-path man-in-the-middle (MitM) attacker who can directly see and modify legitimate readings.} {Such attacks are more common today. However, many such attacks (e.g., through ARP spoofing) can be easily detected/prevented by mainstream network security solutions, e.g., network intrusion detection systems or dynamic ARP inspection by Ethernet switches. 
Also, as discussed earlier, fully controlling a sensor is more difficult task for an attacker than making the sensor unavailable. Hence, in our threat model we assume that the attacker cannot gain direct control of the original sensor nor can she be on-path of the real communication to launch a MitM attack, and we focus on detecting such off-path false-data-injection attackers during their hiding phase.} 

{In particular, a concrete end-to-end attack that combines these two steps was demonstrated by researchers from IOActive in Black Hat USA 2017 conference against a nuclear plant%
~\cite{wired, santamarta2017go}.  %
There, the demonstrated attack blocks the real data from the sensors through a denial-of-service attack and sends spoofed data to a nuclear plant's monitoring system through wireless links. %
The researchers demonstrate the attacks using the same model of devices as used in a real-world nuclear plant and they show that such attacks can potentially lead to severe catastrophic consequences. Also, in vehicular networks, often the ultra-sonic distance sensors of autonomous vehicles and drones are attacked by a similar combination of jamming and spoofing \cite{liu2016can, s20216157}.}

{If the attacker does nothing after making a sensor unavailable (i.e., does not send any fake measurements), the absence of the sensor report will be easily detected as a deviation from the expected behavior of a sensor and can lead to an alarm of anomaly.} %

{One may consider the case that an attacker can make a sensor unavailable only when the attacker is ready to launch the actual attack campaign (i.e., only at the planned time when she actually wants to cause damage). This will minimize the period that the attacker needs to impersonate the sensors and hide away from the detection mechanism proposed in this work. However, there are important reasons why the attacker needs to crash a sensor early.}

{One practical reason is that,  
the best time for an attacker to attack a sensor
is during the system maintenance/upgrading period when more ad-hoc traffic is ongoing or when the attacker has easier access to the system (either physically or remotely). On the other hand, the best time for launching an actual attack campaign to maximize the impact could be during peak hours or other critical moments (e.g., during the holiday season or during critical moments in a war). Some of the attacks to make the original sensors unavailable can only be done when the attacker has physical access to the system (e.g. by physically damaging the sensor like in \cite{lat}), and that time might not be ideal for launching an attack. Also, when the attacker wants to launch a large-scale attack, it may need to prepare these for a prolonged period of time.} 

{Another important reason for the attacker to crash the sensor earlier than the actual attack campaign is because
prior to launching the attack, the attacker may not know if she will be successful or not. For example, an attacker might try to exploit some known vulnerability and is not sure if the targeted device has the latest security patch. In such a scenario, the attacker would need to make an attempt in order to confirm whether it can exploit such a vulnerability. This likely cannot be done at the last minute of the actual attack campaign. Once the sensor crashes, the attacker would not be able to bring the sensor back, hence, to evade detection, the attacker needs to impersonate that sensor by sending out fake measurements. %
}

{Last but not least, for many ICS, especially those equipped with anomaly detection capability, the attacker needs to use the so-called ``frog boiling attack technique''~\cite{10.5555/3327345.3327509, DBLP:journals/corr/abs-2002-02741, 10.1145/1639562.1639592} during her attack campaign, where the attacker tries to gradually poison the system by sending false measurements that are small deviations from the real ones. By gradually doing this, the attacker steers the system state to a state of its liking without triggering alarms. These attacks need to inject a large number of fake data during a ``hidden'' phase, instead of directly sending a fake data with large deviation, as the latter is more likely to trigger an anomaly detection. %
This is another reason for an attack to require a prolonged ``hiding'' period where our defence mechanism will be useful to provide early detection.}

To make the attacker as strong as possible, we assume the attacker can observe any sensor for a significantly long duration, before taking control of the sensor. As such, the attacker is assumed to have all historical data of the system, from which it can gain complete knowledge about the physical system. 
However, the attacker is unaware of the shared secret between a sensor and the defender.

Additionally, for the scope of this work, while trying to stay hidden,  an adversary can only modify sensor output of the compromised sensor and does not change control signals.

\section{Why existing solutions are not desirable}

\noindent Before presenting our solutions, we first discuss some straightforward approaches and explain why they are not desirable.

\para{Using standard encryption or authentication scheme} 
\noindent
One straightforward solution would be to use some secret key to encrypt the sensor's messages, or to %
add a message authentication code to the sensor's readings.
Overall, such techniques inherently require all affected devices in the ICS to be upgraded to accommodate the introduced changes, which might be rather costly.
Also, if the sensor's reading is consumed by multiple devices in a broadcast or multicast group, sharing a single key exposes it to a large attack surface. Any compromised member in the group can impersonate that sensor. Using asymmetric keys can mitigate this risk, but incurs higher computation overhead on both the sensor and the other devices.
In contrast, in our design the secret key is just shared between the distortion source and the defender. Our design does not require changes to the original sender and receiver devices and also imposes smaller attack surface. 

\para{Simpler ways to show the possession of shared secret by the sensor}
\noindent We see that introducing the standard encryption and authentication schemes is against our design goal. However, there are still simpler ways to let the sensors use the key to authenticate itself to the defender.
One approach is to just send the secret key $k_i$ out of band to the defender. Another approach is to let the sensor double its sending rate and always send readings in a pair, where the first value in the pair is the original reading $d_i$, and the following one is the distorted reading $d'_i$.
There are a few problems with these simpler approaches for a sensor to show to the defender that it possesses the shared secret: (a) this cannot be achieved if the distortion is introduced through the physical mean, as the sensor will still send out in its original sending rate and cannot send out separate message streams containing the secret. (b) even when the distortion is introduced digitally, the additional secret stream from the sensor or the doubling of the sending rate may cause unexpected effects on legacy devices that depend on the sensor readings. If the sending of the keys is made totally independent from the sending of the sensor readings (hence reducing the chance of affecting the legacy devices), it may be possible for an attacker to crash the software process that transmits the sensor readings without crashing the process that transmits the one-time secret. 

\section{Detection Using Digital Micro-Distortion}
When digital manipulation of the sensor data through a secured device is possible, hidden attackers can be detected by simply updating the least significant bit (LSB) of a sensor reading using the secret key. Here, for the sensor reading at time slot $i$, the digital distorter just needs to simply rewrite the sensor reading's LSB to be the $i^{th}$ bit of the secret key.
For a perfect one-time pad where the attacker cannot guess the current bit of the secret key even when it is aware of all the previous bits, this mechanism can lead to fast and accurate attack detection.

\begin{figure*}[t!]
  \centering
   \subfloat[]{ \includegraphics[scale=0.26]{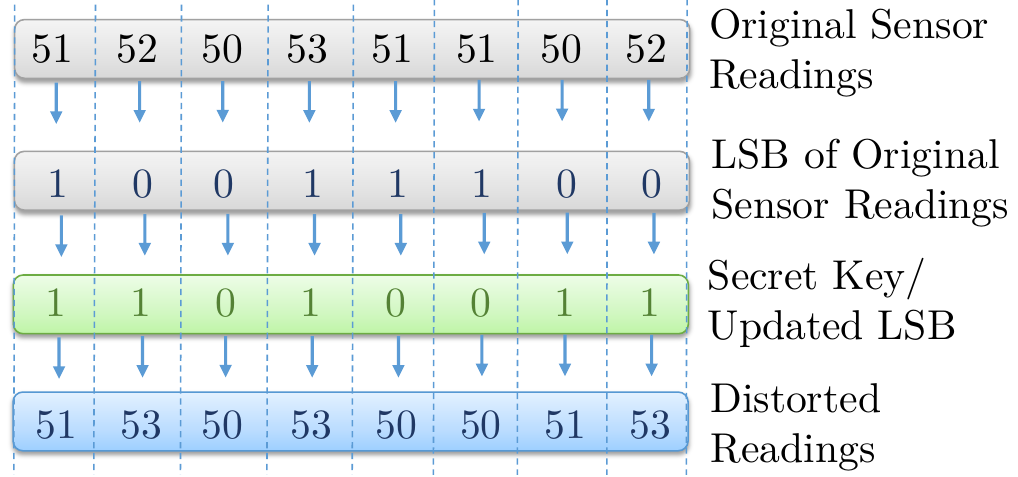}} \hfill
    \subfloat[]{\includegraphics[scale=0.26]{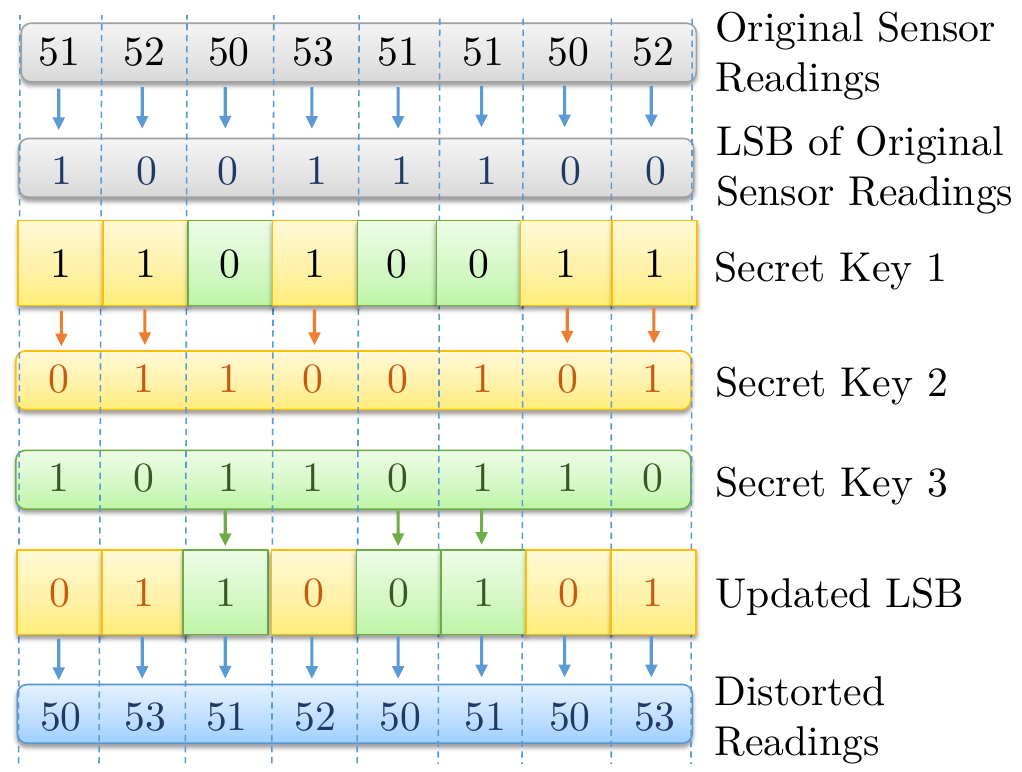}} \hfill
    \subfloat[]{\includegraphics[scale=0.26]{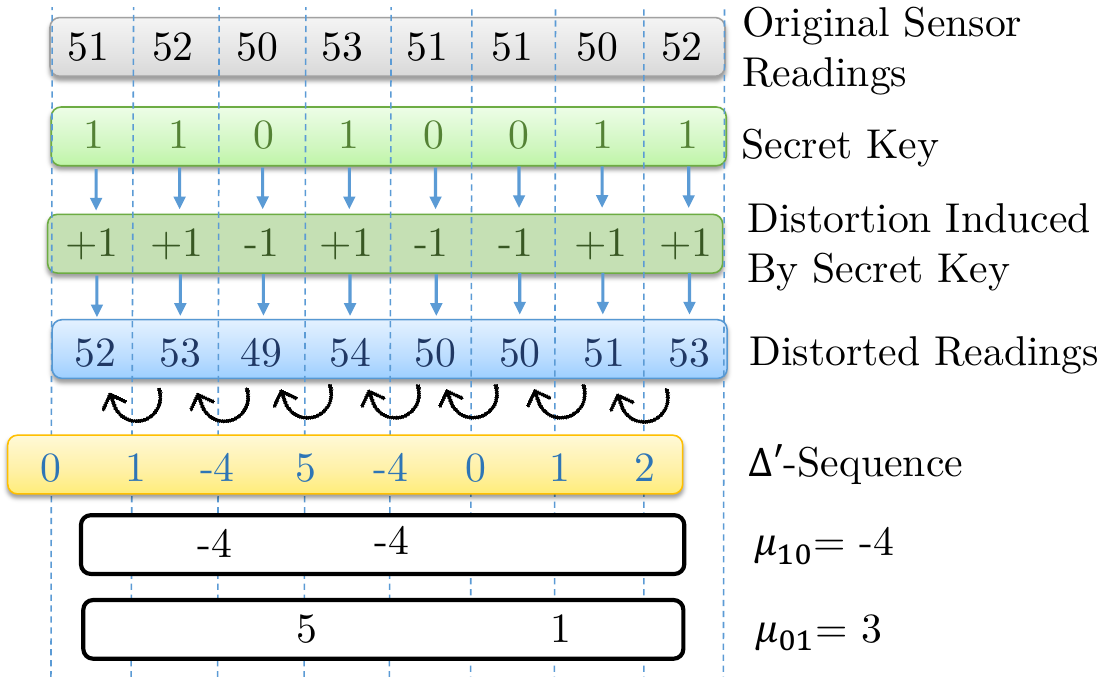}}
  \caption{\small(a) Shows the addition of digital distortion through updating the LSB using a single secret key. (b) Shows the two-layer defense using three secret keys. In each time-slot, key 1 determines whether key 2 or key 3 is used for updating the LSB. (c) Shows the addition of physical distortion while illustrating the different notations used.}
  \vspace{-4mm}
  \label{fig:distortion}
\end{figure*}

In practice, however, given that the one-time pad is imperfect, a potential drawback of this approach is that the secret key can now be observed directly by an attacker (i.e., she just needs to extract the LSB of the sensor readings). 
In case of a weak one-time pad (e.g., that is generated with non-truly-random data or with a weak generation algorithm or when padding parts are re-used), the attacker can use this historical data to break the system's security (see \cite{deavours1985machine} for such known plain-text attacks). To mitigate such risk, %
we propose a two-layer security mechanism that uses three independent streams of secret keys ($sk1$, $sk2$, and $sk3$). Here, the role of the first secret key ($sk1$) is essentially to randomly mix the second and third secret keys ($sk2$  and $sk3$) in a way that makes it more difficult for any attacker relying on pattern determination to predict any of the keys. The secret key at the first stage ($sk1$) is used to dictate the timing of the use of the other secret keys. That is, in the second stage, for each instance either $sk2$ or $sk3$ is used as determined by $sk1$. For example, consider that the sensor reading's LSB to be digitally updated to the $i^{th}$ bit of $sk2$, only at the instances when the $i^{th}$ bit of $sk1$ is $1$ (w.l.o.g.). Similarly, at the instances when the $i^{th}$ bit of $sk1$ is $0$ (w.l.o.g.), the sensor reading's LSB is updated to the $i^{th}$ bit of $sk3$. This two-layer defense provides different sources of entropy, leading to a non-linear combination of the secret keys, thereby increasing the difficulty required to decipher them. Even when the LSB of the actual sensor reading is known to the attacker and the attacker can see the associated distorted outgoing sensor readings, it would be difficult for the attacker to recover any of the three secret key streams. %
Figures \ref{fig:distortion} (a) and (b) show examples of digital insertion of distortions for the simple and the two-layered scheme, respectively.

For this case, we assume reliable data transmission between the distortion source and the detector. (Detecting and correcting transmission errors is the standard practice in communication networks used in ICS). Hence, we can minimize the magnitude of the micro-distortion by manipulating only the LSB in the sensor reading. This %
can lead to extremely fast attack detection with an error rate that reduces exponentially with the number of trials. To successfully evade detection, the attacker has to correctly predict the secret key bit used in each time-slot.
So, the probability of false negative outcome over $t$ time-slots is given by
\[\Prob{FN} = \left( \Prob{\text{Correct guess by the attacker}} \right)^t\]
Note that $\Prob{FN}$ quickly comes down to $0$ as $t$ increases, for as long as $\Prob{\text{Correct guess by the attacker}}$ is $<1$. 
Specifically, for our designed two-layer scheme, 
the probability of the attacker correctly guessing any bit is closer to $0.5$. For as low as $20$ sensor readings, the probability that an attacker can successfully guess all $20$ bits correctly is lower than $10^{-6}$.

Though detection can be fast, as discussed earlier, the addition of a digital distorter can result in additional attack surface given the bump-in-the-wire setting. It might make it easier for a remote attacker that can just compromise the digital distortion devise to get hold of the secret key, which would then invalidate the basic premise of the detection mechanism. In comparison, a physical distortion device can be kept isolated (i.e., with an air gap) from the communication network, making it almost impossible for a remote attacker to gain control of it. 
Another limitation of using such a digital distortion device is the cost consideration. %
The cost involved here consists of the cost to add the bump-in-the-wire hardware or the cost to update the sensor's firmware. More importantly, such changes require vendor support, may void the warranty, and require a costly re-certification.
The expense of doing so can often go against our design goal of getting a low-cost solution. As such, in the following sections, we will discuss the physical addition of micro-distortion.

\section{Detection Using Physical Micro-distortion} \label{Sec:Detection_Algo}

\subsection{Adding Micro Distortions Physically} %
Different from the digital means of introducing micro-distortion, in this approach, we will deploy a micro-actuator that will physically introduce the micro changes to the underlying
system, so that the original sensor will pick that up in
its reading. For example, if the sensor is measuring the power consumption of a system, we introduce a small programmable load that can be turned on or off based on the secret key.
Specifically, for each sensor, we determine the micro-distortion $\epsilon$ value based on the magnitude of the sensor readings (e.g., $< 0.5\%$ of the sensor's operating range).  For each sensor reading $d_i$,  if the corresponding key-value $k_i$ is $1$, then the sensor's reading is distorted by adding a value of $\epsilon$ to the reading; else is distorted by subtracting a value of $\epsilon$ from the reading.  Note that, adding distortions in this form maintains a zero mean distortion in the long run. 
In other words, 
given $k_i$ (the secret key for a time slot $i$),
the sensor's original data reading $d_i$ is changed to 
$d'_i = d_i + (2k_i-1) \epsilon$. Specifically, an increment or decrement by $\epsilon$ based on the value of $k_i$.

{Do note that our design for physically adding micro-distortion is inherently robust to analog noise arising due to device imperfections, measurement errors, or environmental factors,  given that our technique takes multiple samples to detect the presence or absence of the micro-distortions. Consider the micro-distortion as the signal from an authenticated sensor that our detector seeks to decode. The noise (be it natural variation in the original system, the measurement errors, or the device imperfections) can be handled by looking at multiple samples to achieve a high decoding accuracy despite of varying signal-to-noise ratio. In the more challenging case, it will take a larger number of readings and a longer time to detect a potential attack in highly noisy systems. In fact, we have run our proposed method using noisy real-world data and our experimental results demonstrate the robustness of our approach under such real-world analog noise. }

\subsection{Magnitude of the Micro-distortion}

The magnitude of the micro distortion ($\epsilon$) can be selected based on the following two key considerations.

Firstly and most importantly, it should not cause a noticeable disturbance to the system and should be tolerable by the system. Hence, the chosen value should not be greater than the natural variation of the targeted metric or the natural noise level --- e.g., as in our experiments, the power supply and demand in a power system can naturally change at a level above 0.5\%, making our chosen magnitude of 0.5\% and 0.25\% in the conducted experiments small enough to not cause any disturbance. In comparison, for measurements like the frequency of the power system, which needs to be much more stable, the allowed distortion magnitude would need to be significantly smaller.

Under the above hard constraint, the second consideration is the trade-off between the attack detection delay and the magnitude of distortion introduced into the system. 
One can consider the micro-distortion as the signal from an authenticated sensor that our detector seeks to decode. The natural variation in the original system can be considered as the noise that can cause difficulty for the detector to decode the signal. Hence, a higher magnitude of the micro-distortion increases the signal-to-noise ratio, and hence makes it faster for the defender to determine the presence or absence of potential hidden attackers. In comparison, a lower magnitude of the micro-distortion can lead to a longer detection delay as the defender needs to collect more samples to determine whether the micro-distortion is present or not; however, a lower magnitude distortion makes it easily tolerable by the system.

For the digital micro-distortion case, as we assume reliable data transmission between the distortion source and the detector, we minimize the magnitude of the micro-distortion by manipulating only the %
LSB in the sensor reading.

\subsection{A Straw Man Detection Scheme: Simple Mean Difference}
If one considers a perfectly stable noiseless system where the sensor readings remain constant for the detection period, %
it is easy to see that our approach can detect a compromised sensor extremely fast, by simply letting the defender check for the pattern of the distortion based on the same secret.
An attacker without the knowledge of the secret would not be able to replicate the pattern, hence cannot bypass the detection. In fact, if the attacker just makes a random guess, in each slot, it has a $50\%$ chance of guessing wrongly and therefore being detected. As discussed in the digital distortion section, the probability that the attacker can remain undetected after $20$ slots is as low as $0.5^{20} < 10^{-6}$. However,  under normal functioning of an ICS, the sensor readings are subject to state changes in the ICS along with possible noise. With the possible sensor readings spanning a much wider (e.g., 200x) range of a micro-distortion, the micro-distortion becomes a negligible signal that easily gets overwhelmed by the magnitude of the actual sensor readings.

\para{Simple Mean Difference} 
By the ``law of large numbers" \cite{dekking2005modern},
the mean value of a large number of observations made for a random variable 
approaches the random variable's expected value as more observations are taken. 
Consider the set of readings distorted as $d'_i=d_i+\epsilon$ as set $S_1$ (i.e., with corresponding secret key $k_i=1$), and the remaining set as $S_0$, i.e., with $k_i=0$ and corresponding reading distorted as $d'_i=d_i-\epsilon$). Since we select the two sets from all the readings over a time window based on the random secret, we can view the original readings $d_i$ from both sets as observations drawn from the same distribution. Hence, with a larger number of observations, the difference between the mean value of all distorted readings from set $S_1$ and the mean value of all distorted readings from set $S_0$ should approach $2\epsilon$. If the attacker does not know the secret key, it cannot introduce any statistical difference between these two sets. As such, the difference of the mean value for $S_0$ and $S_1$ should approach $0$ when under attack. 

This shows that the detection can eventually be achieved if the detector can examine a sufficiently large number of samples. The issue of simply relying on the law of large numbers, however, is that the detection can be rather slow. For example, if the actual readings are drawn uniformly randomly and independently from all the possible range of values, and when the micro-distortion $\epsilon$ equals to $0.5\%$ of the possible range, our evaluation result shows that it requires more than 80,000 samples in order to reduce both the false positive and false negative rate below $1\%$. Similarly, it needs more than 140,000 samples in order to further reduce that to below $0.1\%$. Even if the sensor reading is sent every second, this translates to almost one whole day for achieving $1\%$ false positive/negative rate, and nearly $40$ hours of readings for $0.1\%$ false positive/negative rate. If the sensor reading is sent only every minute, it will further inflate the required detection delay to around $2$ and $3.5$ months, respectively.

\subsection{Our Design: $\Delta$ based Mean Difference}

Fortunately, in real-world ICS, the sensor readings present some good statistical properties that allow much faster detection of such attacks. Specifically, in many ICS (and power grid in particular), physical properties being sensed (e.g., the amount of power being generated or consumed) tend to change in a gradual manner (i.e., with small differences between consecutive time slots) in a significant fraction of time.
Based on that, we propose a %
detection algorithm, where 
instead of comparing the difference of the mean of the sensor readings from sets $S_1$ and $S_0$ (as defined earlier), we look at the change in sensor reading between consecutive time slots, which we refer to as $\Delta$.

\para {$\Delta$-sequence creation} Given the distorted sensor reading sequence $d'_1, ....,  d'_n$ and the secret key sequence $k_1, ..., k_n$, 
we define the $\Delta$ sequence as $\Delta_1, ..., \Delta_{n-1}$ and $\Delta'$ sequence as $\Delta'_1, ..., \Delta'_{n-1}$, where
\[\Delta_i = d_{i+1} - d_i \text{ and } \Delta'_i = d'_{i+1} - d'_i\]
While $\Delta_i$ gives the difference between the original sensor readings in consecutive time slots, $\Delta'_i$  gives the difference between the distorted sensor readings in consecutive time slots.

\para{Data partitioning step} %
We define set $S_{01}$ as the collection of all moments $i$ such that $k_i=0$ and $k_{i+1}=1$ and we define set $S_{10}$ as the collection of all moments $i$ such that $k_i=1$ and $k_{i+1}=0$. We define set $S_{00}$ and set $S_{11}$ similarly.
It could be seen that for an $i$ that belongs to different sets, the relationship between the corresponding $\Delta_i$ and $\Delta'_i$ is different. Specifically, for $i\in S_{00}$ or $i\in S_{11}$, since the same distortion is applied to both $d'_i$ and $d'_{i+1}$, we can see that $\Delta'_i = \Delta_i$. On the other hand, $\Delta'_i = \Delta_i + 2\epsilon$ for $i\in S_{01}$, while $\Delta'_i = \Delta_i - 2\epsilon$ for $i\in S_{10}$. 
See Figure \ref{fig:distortion}(c) for an example of physical addition of micro-distortion.

Since each of the random key $k_i$ is drawn with equal probability from $0$ and $1$ in an independent manner, it is easy to see that a moment $i$ (in regard to $\Delta$ and $\Delta'$ sequence) has an equal probability of falling into one of the four sets $S_{01}$, $S_{10}$, $S_{00}$, and $S_{10}$. As the value in the $\Delta$ sequence does not depend on the value of the secret key sequence, we have:
\begin{eqnarray}
\Expt{avg(\Delta_i | i \in S_{01})}=\Expt{avg(\Delta_i | i \in S_{10})}\nonumber\\
=\Expt{avg(\Delta_i | i \in S_{00})}=\Expt{avg(\Delta_i | i \in S_{11})}\nonumber
\end{eqnarray}
$$\text{Consider the gauge }x=avg(\Delta'_i | i \in S_{01}) - avg(\Delta'_i | i \in S_{10}) $$
We have:
\begin{eqnarray}
& & \Expt{x}\nonumber = \Expt{avg(\Delta'_i | i \in S_{01}) - avg(\Delta'_i | i \in S_{10})} \nonumber\\
& = & \Expt{avg((\Delta_i + 2\epsilon) | i \in S_{01}) - avg((\Delta_i - 2\epsilon) | i \in S_{10}})  \nonumber\\
& = & 4 \epsilon + 
\Expt{avg(\Delta_i | i \in S_{01})} - \Expt{avg(\Delta_i | i \in S_{10})}  \nonumber \\ & = &4 \epsilon \nonumber
\end{eqnarray}

\begin{figure*}[]
	\centering
	\subfloat[Original data]{\includegraphics[width=0.27\textwidth]{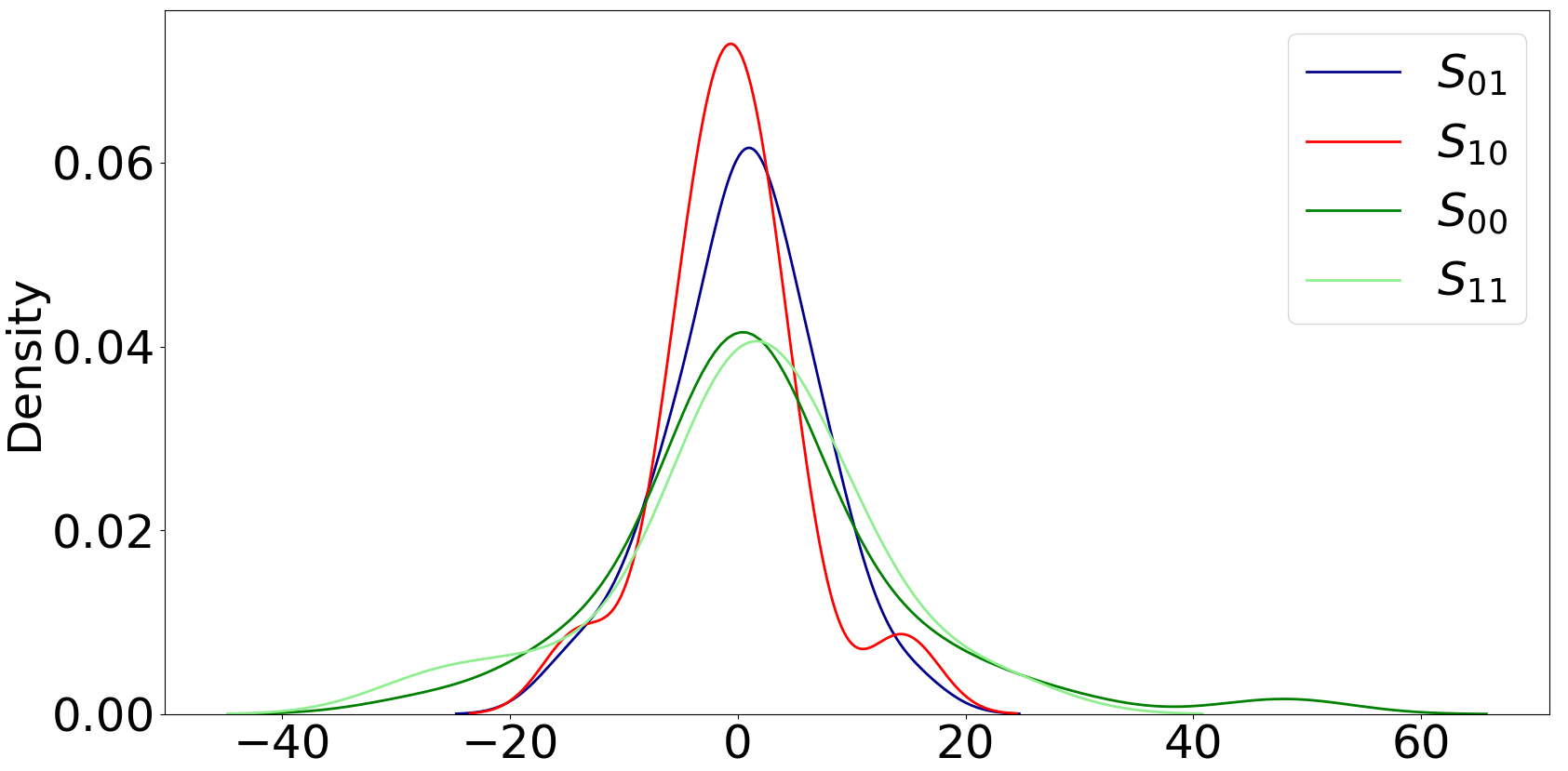}}
	\hspace{0.2em}
	\subfloat[With added micro-distortion]{\includegraphics[width=0.27\textwidth]{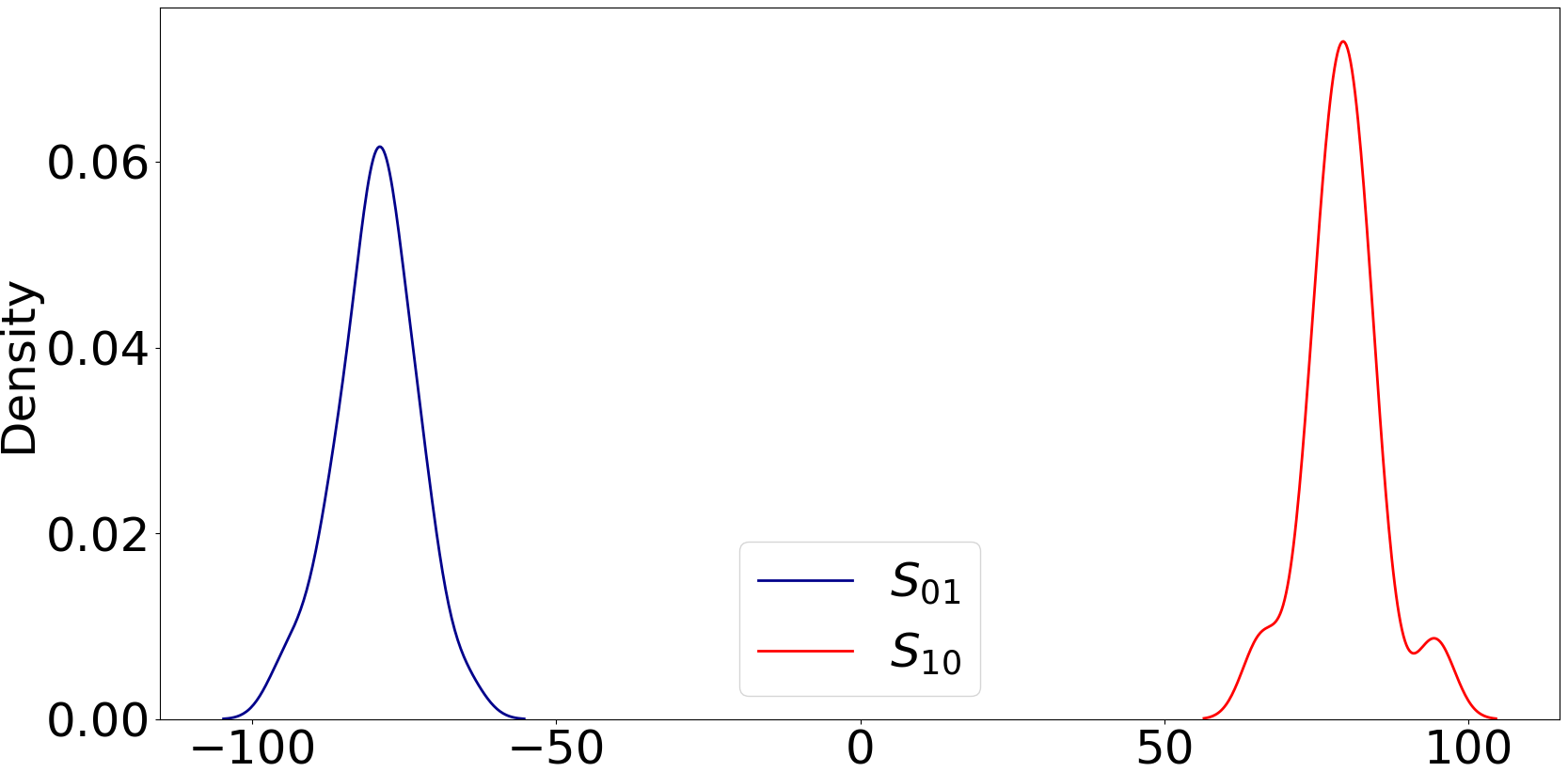}}
	\hspace{0.2em}
	\subfloat[Random distortion attack]{\includegraphics[width=0.27\textwidth]{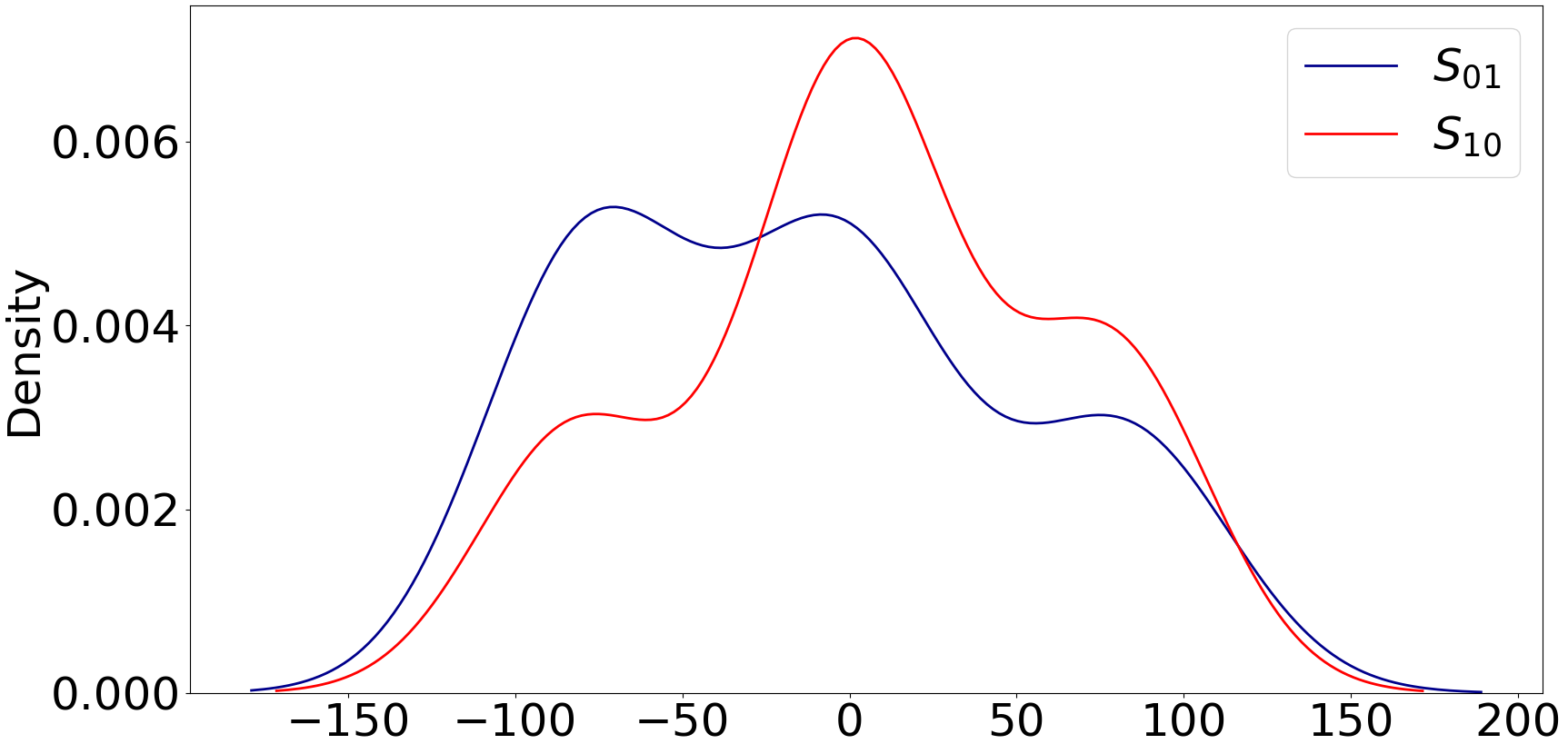}}
	\caption{\small An example of a probability density distribution of the $\Delta_i$ values %
	for a case taken from the Smart Grid Meter dataset, where sequences $S_{01}$ and $S_{10}$ can be clearly distinguished after the addition of micro-distortion. Fig. 3(a) shows the probability density of the $\Delta_i$ values obtained from the original data for sequences $S_{00}$, $S_{11}$, $S_{01}$, and $S_{10}$. Fig. 3(b) shows the probability density of the $\Delta_i$ values for sequences $S_{01}$ and $S_{10}$ after adding the micro-distortion. Fig. 3(c) shows the probability density of the $\Delta_i$ values for sequences $S_{01}$ and $S_{10}$ when an attacker randomly injects the micro-distortions. \vspace{-3mm}}
	\label{fig:datasets}
\end{figure*} 

\begin{figure*}[]
	\centering
	\subfloat[Original data]{\includegraphics[width=0.26\textwidth]{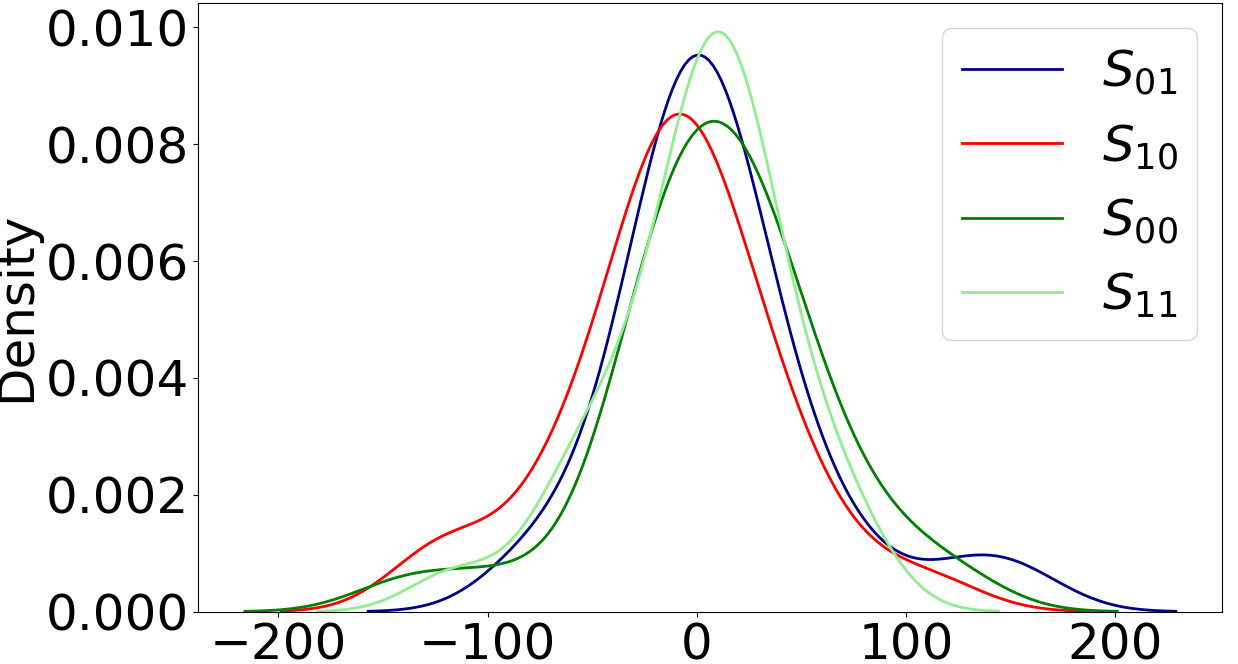}}
	\hspace{0.2em}
	\subfloat[With added micro-distortion]{\includegraphics[width=0.26\textwidth]{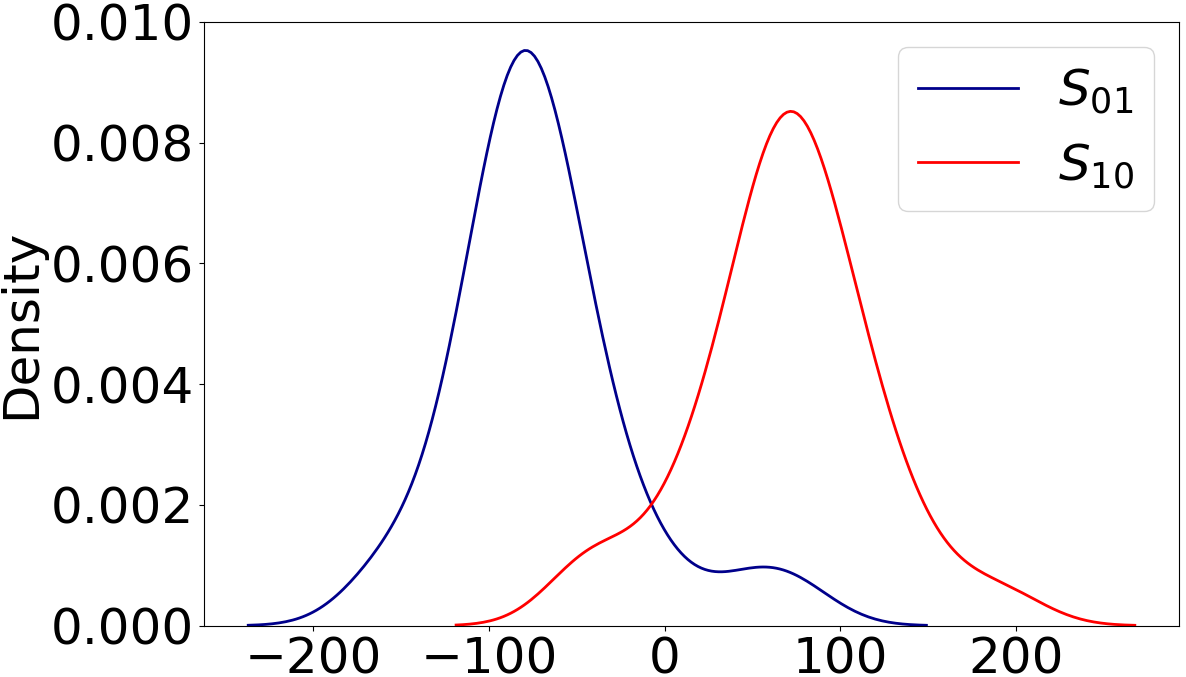}}
	\hspace{0.2em}
	\subfloat[Random distortion attack]{\includegraphics[width=0.26\textwidth]{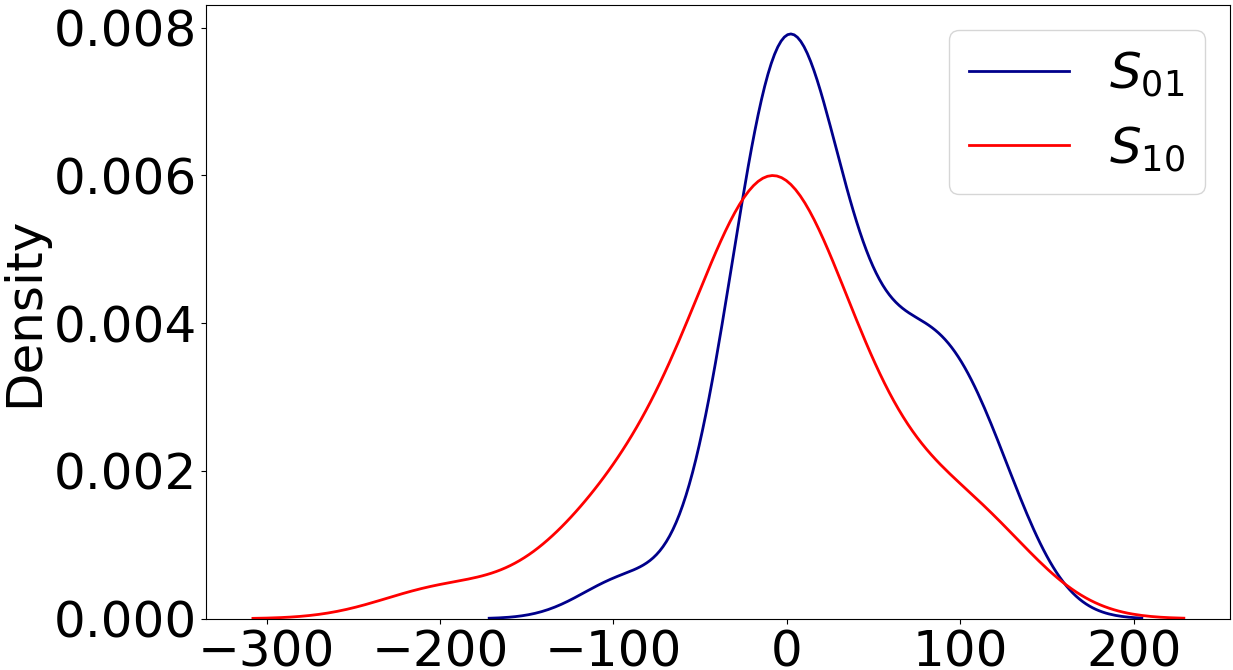}}
	\caption{\small Another example of a probability density distribution of the $\Delta_i$ values for another case taken from the Smart Grid Meter dataset, where the chosen sample size $n$ is insufficient to make a clear distinction between sequences $S_{01}$ and $S_{10}$ after the injection of micro-distortion (due to the overlapping region in Fig 4(b)). %
	\vspace{-3mm}}
	\label{fig:datasets1}
\end{figure*} 

An example of this separation can be seen in Figures \ref{fig:datasets}(b) and \ref{fig:datasets1}(b). In Figures \ref{fig:datasets}(a) and \ref{fig:datasets1}(a) we see that in the original data, the probability density function of the $\Delta_i$ values of the four sets are somewhat similar (given that the sets are chosen randomly). In Figure \ref{fig:datasets}(b), once the micro-distortions are added, a clear separation of the $\Delta_i$ mean values can be seen between the sets $S_{01}$ and $S_{10}$. However, in Figure \ref{fig:datasets1}(b), we see that the number of samples considered ($n$) was insufficient to introduce a clear separation of the $\Delta_i$ mean values, which could possibly lead to incorrect attack detection. In this case, considering data from a longer duration, i.e., a larger sample size, would increase accuracy. Lastly, in Figures \ref{fig:datasets}(c) and \ref{fig:datasets1}(c), we see that if an attacker were to randomly inject micro-distortions, without knowing the sets $S_{01}$ and $S_{10}$, then the attacker will not be able to introduce any significant statistical difference between sets $S_{01}$ and $S_{10}$, leading to detection.

\para{Detection Condition} As shown, if we calculate the difference between the mean of all $\Delta'_i$ in set $S_{01}$ and that in set $S_{10}$ as $x$, the expected value of $x$ should be $4\epsilon$ when there is no attack. In comparison, in case of an attack, as we assume that the attacker does not know the value of $k_i$, it cannot introduce any significant statistical difference between the set $S_{01}$ and $S_{10}$. As a result, in this case, the expected value of $x$ should approach $0$. In other words, we can use the expected value of $x$ to differentiate between the attack and non-attack cases.

In particular, we calculate the $\Delta$ mean difference $\mu_{01} = avg(\Delta'_i | i \in S_{01})$ and $\mu_{10} = avg(\Delta'_i | i \in S_{10})$. Thereafter, we check whether {$\mu_{01} - \mu_{10} \in \{2\epsilon, 6\epsilon\}$} (as the difference concentrates near the expected value of $4\epsilon$) and raise alarm to detect attack if the condition is not satisfied.

While this seems similar to using the expected difference of the mean value between $S_0$ and $S_1$ (the procedure, we refer to as `Simple Mean Difference'), the benefit of calculating using $\Delta'$ sequence is that, for many ICS, the absolute value of the elements in the $\Delta'$ sequence can be significantly lower than those in the distorted reading sequence (i.e., the $d'$ sequence). This is because in many ICS (including many power grid systems), while a particular physical measurement can have readings that span a large range (e.g., the peak power generation or consumption in an energy system can be $10\times$ or even $100\times$ bigger than its non-peak period), it turns to change gradually at most times. Thereby making the distribution of value in $\Delta$ and $\Delta'$ sequence concentrate  more heavily towards smaller values, i.e., the variance of the corresponding sequence is much smaller. The smaller values, in turn, make it possible to use a small number of samples to approach a given (small) neighborhood of the expected value with a higher probability.

\para{Filtration Step} Though the above steps provide a complete detection algorithm, %
in many systems, although most of the changes across two consecutive time slots have low magnitude, there can be some high-magnitude changes from time to time. For example, when a household turns on a heater, its power demand can increase dramatically, although most of the time, the power demand in the household only has small changes.
Filtering out some high $\Delta'_i$ values %
(considering absolute values of $\Delta'_i$) that can cause high variance in the $\Delta'$ sequence can often result in significant improvements. %
For systems with intermittent large abrupt changes (like in power-grid systems), %
even though this filtration would reduce the sample size, it would significantly bring down the variation as well, making it much easier for attack detection while also improving the accuracy. This fact becomes more evident from our experiments on the smart grid meter dataset and the SWaT dataset (see Sections \ref{sec:house} and \ref{sec:swat}). The SWaT dataset considers the change in water-levels in different tanks and therefore does not have any large abrupt changes. In contrast, the smart grid meter dataset measures the power usage of a particular household, where abrupt and large changes occur due to switching on (or switching off) of an appliance. We observe that having a filtration step significantly improves the detection time and accuracy for the smart grid dataset, whereby making almost no difference for the SWaT dataset. Consequently, the $|\Delta'_i|$ readings that are greater than a particular threshold $\Delta_{th}$ are removed from consideration, where $||$ represents the absolute value function. %
$\Delta_{th}$ is based on the past (correct) operation of the sensor and is determined in %
a way that the number of $\Delta'_i$ readings removed is not too much for the time duration considered, i.e., $n$.

However, an attacker might take advantage of such a filtration procedure by introducing high noise to faked sensor outputs which would likely result in most of the noisy data being filtered out, thus delaying the detection of the attacker. Such attackers can be checked by choosing another threshold $m$ (based on the $\Delta_{th}$ and system under consideration) which ensures that there is always sufficient $\Delta'_i$'s that get through even after the filtration step when there is no attack. %

See Algorithm \ref{alg:detection} for pseudocode for the `filtered $\Delta$ mean difference algorithm'. In the absence of the filtration step, we refer to the algorithm as `$\Delta$ mean difference algorithm'.

\SetNlSty{textbf}{}{:}%
\IncMargin{.2em}%
\setlength{\textfloatsep}{0pt}
\begingroup
\LinesNumberedHidden
\begin{algorithm}[!t]
\DontPrintSemicolon
\caption{Filtered $\Delta$ Mean Difference Algorithm.}
\label{alg:detection}
 \KwIn{Given a time, denote the latest $n$ time slots as time slot $1$ to $n$, with corresponding pre-shared secret $\{k_1, k_2, \dots, k_n\}$ and the sensor readings $\{d'_1, d'_2, \dots, d'_n\}$ (which is supposed to be distorted according to the secret when there is no attack.}
 \KwOut{Raise an alarm if detecting the presence of an attacker in the given time interval.}
\Numberline \textbf{$\Delta'$-sequence creation: } 
Create $\Delta'$-sequence such that 
  $\Delta'_i = d'_{i+1} - d'_{i},  \forall i=1, \ldots, n-1$\\
\Numberline\textbf{Filtration Step: } All rows for which $|\Delta'_i| > \Delta_{th}$ are removed from consideration. \\
\Numberline\textbf{Data Partitioning Step: } Identify set $S_{01}$ as all time indices $i$ such that $k_i=0$ and $k_{i+1}=1$. Identify set $S_{01}$ as all time indices $i$ such that $k_i=1$ and $k_{i+1}=0$.\\ 
\Numberline \If{size of remaining $\Delta'_i$ with $i \in S_{01} \cup S_{10}$ is less than a threshold $m$}
								{Raise alarm to detect attack.}
\Numberline \Else{Calculate $\mu_{01} = avg(\Delta'_i | i \in S_{01})$ and $\mu_{10} = avg(\Delta'_i | i \in S_{10})$. \\
\If{$\mu_{01} - \mu_{10} \notin \{2\epsilon, 6\epsilon\}$}{Raise alarm to detect attack.} }
\end{algorithm}
\endgroup

\begin{table}[!h]
\small
\begin{center}
\resizebox{0.7\columnwidth}{!}{
\begin{tabular}{|c|c|c|}
\hline
& Power (W) & $\Delta$-sequence (W)\\ \hline
Maximum Value & 17206.0  &  14351.0\\ \hline
Minimum Value &  225.0  &  0.0\\ \hline
Average Value & 1104.0733  &  13.611\\ \hline
Median Value &  775.0  &  3.0\\ \hline
\end{tabular}}
\end{center}
\caption{\small Table showing the power usage data statistics for House 1 and its associated $\Delta$-sequence.}
\label{tab:house_data}
\end{table}

As a by-product benefit of our solution, observe that now the one-time pad is not sent in its clear form through any communication between a sensor and the defender. There always exists the possibility of misinterpretation of a natural change of the sensor reading as a micro-distortion to any wiretapper or observer. So, even when the algorithm generating the one-time pad is weak, %
our use of the one-time pad hidden within the naturally occurring changes will make it harder for the attacker to exploit the weakness.

\begin{table}[ht!]

\centering
\resizebox{\columnwidth}{!}{
\begin{tabular}{|c|ccc|ccc|ccc|}
\hline
\multirow{3}{*}{n} & \multicolumn{3}{c|}{\begin{tabular}[c]{@{}c@{}}Simple Mean\\ Difference\end{tabular}}                                    & \multicolumn{3}{c|}{\begin{tabular}[c]{@{}c@{}}$\Delta$ Mean\\ Difference\end{tabular}}                                           & \multicolumn{3}{c|}{\begin{tabular}[c]{@{}c@{}}Filtered $\Delta$\\ Mean Difference\end{tabular}}                                  \\ \cline{2-10} 
                   & \multicolumn{1}{c|}{\multirow{2}{*}{\begin{tabular}[c]{@{}c@{}}FP\\ (\%)\end{tabular}}} & \multicolumn{2}{c|}{FN (\%)}        & \multicolumn{1}{c|}{\multirow{2}{*}{\begin{tabular}[c]{@{}c@{}}FP\\ (\%)\end{tabular}}} & \multicolumn{2}{c|}{FN (\%)}        & \multicolumn{1}{c|}{\multirow{2}{*}{\begin{tabular}[c]{@{}c@{}}FP\\ (\%)\end{tabular}}} & \multicolumn{2}{c|}{FN (\%)}        \\ \cline{3-4} \cline{6-7} \cline{9-10} 
                   & \multicolumn{1}{c|}{}                                                                 & \multicolumn{1}{c|}{EDA}  & RDA  & \multicolumn{1}{c|}{}                                                                 & \multicolumn{1}{c|}{EDA}  & RDA  & \multicolumn{1}{c|}{}                                                                 & \multicolumn{1}{c|}{EDA}  & RDA  \\ \hline
30                 & \multicolumn{1}{c|}{6.95}                                                             & \multicolumn{1}{c|}{1.69} & 2.53 & \multicolumn{1}{c|}{4.22}                                                             & \multicolumn{1}{c|}{0.97} & 3.94 & \multicolumn{1}{c|}{0.19}                                                             & \multicolumn{1}{c|}{0.00} & 0.95 \\ \hline
60                 & \multicolumn{1}{c|}{7.59}                                                             & \multicolumn{1}{c|}{2.73} & 3.05 & \multicolumn{1}{c|}{5.77}                                                             & \multicolumn{1}{c|}{2.27} & 2.73 & \multicolumn{1}{c|}{0.1}                                                              & \multicolumn{1}{c|}{0.00} & 0.46 \\ \hline
90                 & \multicolumn{1}{c|}{6.99}                                                             & \multicolumn{1}{c|}{2.55} & 2.86 & \multicolumn{1}{c|}{5.27}                                                             & \multicolumn{1}{c|}{2.17} & 2.3  & \multicolumn{1}{c|}{0.06}                                                             & \multicolumn{1}{c|}{0.00} & 0.08 \\ \hline
120                & \multicolumn{1}{c|}{7.27}                                                             & \multicolumn{1}{c|}{2.24} & 2.45 & \multicolumn{1}{c|}{4.15}                                                             & \multicolumn{1}{c|}{1.77} & 2.05 & \multicolumn{1}{c|}{0.05}                                                             & \multicolumn{1}{c|}{0.00} & 0.01 \\ \hline
150                & \multicolumn{1}{c|}{6.81}                                                             & \multicolumn{1}{c|}{2.16} & 2.36 & \multicolumn{1}{c|}{4.03}                                                             & \multicolumn{1}{c|}{1.93} & 2.08 & \multicolumn{1}{c|}{0.00}                                                             & \multicolumn{1}{c|}{0.00} & 0.00 \\ \hline
20000              & \multicolumn{1}{c|}{0.89}                                                             & \multicolumn{1}{c|}{0.41} & 0.43 & \multicolumn{1}{c|}{0.00}                                                             & \multicolumn{1}{c|}{0.00} & 0.00 & \multicolumn{1}{c|}{0.00}                                                             & \multicolumn{1}{c|}{0.00} & 0.00 \\ \hline
40000              & \multicolumn{1}{c|}{0.11}                                                             & \multicolumn{1}{c|}{0.00} & 0.00 & \multicolumn{1}{c|}{0.0}                                                              & \multicolumn{1}{c|}{0.00} & 0.00 & \multicolumn{1}{c|}{0.0}                                                              & \multicolumn{1}{c|}{0.00} & 0.00 \\ \hline
\end{tabular}
}
\caption{\small Table showing the false positive and the false negative rate (expressed as percentage) for different $n$ for simple, $\Delta$, and filtered $\Delta$ mean difference algorithms done over 10,000 trials for the smart grid dataset with $\epsilon = 40$ W.  For filtration $\Delta_{th} = 200$ W.}
\label{tab:fp_house}
\end{table}

\section{Experiments and Evaluations} \label{sec:expeval}

In this section, we run our proposed detection algorithm on some real-world datasets and evaluate their performance.  %
In each case, we observe the sensor readings for the duration that it takes to obtain $n$ samples, after which the algorithm outputs the presence/absence of an attacker.

\para{Simulated Attacks} To test the effectiveness of our proposed detection algorithm, we consider two types of attacks. In the first attack, the attacker predicts the exact sensor readings (without the addition of $\epsilon$) and uses that as the faked sensor readings to impersonate the compromised sensor. We refer to this attack as the ``Exact Duplication Attack" (EDA). In the second form of attack, to the exact prediction, the attacker randomly injects micro-distortions, i.e., adds or subtracts $\epsilon$ randomly. We refer to this attack as the ``Random Distortion Attack" (RDA). %

\begin{figure}[htb!]
\centering
\includegraphics[width=0.6\linewidth]{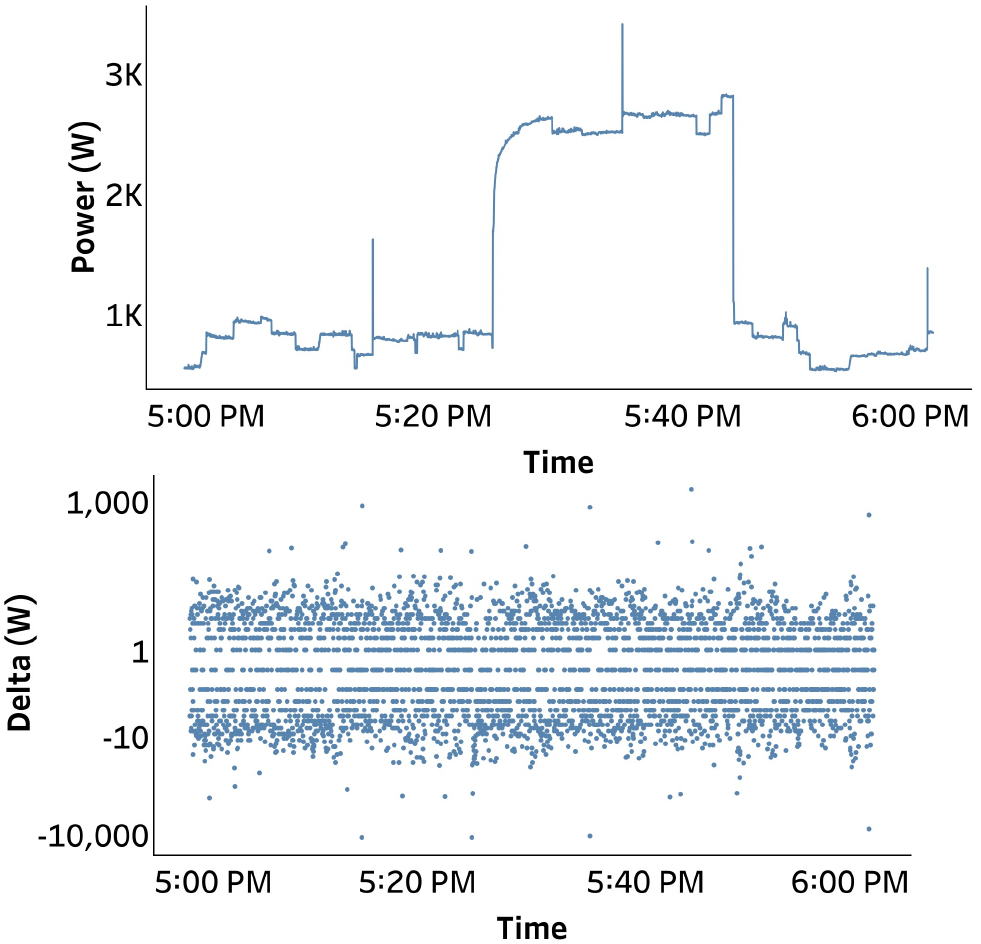}

\caption{ Figure depicting typical fluctuations in smart grid dataset. The variation in $\Delta$ is plotted in $\log$ scale.}
\label{fig:house}
\end{figure}

\addtolength{\topmargin}{0.05in}
\subsection{Attack Detection in Smart Grid Meter Dataset} \label{sec:house}
We make use of the publicly available ``Rainforest Automation Energy Dataset for Smart Grid Meter Data" \cite{DVNZJW4LC2017} that contains 1 Hz data %
from two residential houses over a period of 72 days for House 1 and  59 days for House 2.  %
Overall,  this dataset contains over 11.3 million power readings.  %

We consider the total power usage data for House 1 given by the sensor `mains'. %
The data statistics and the statistics of its associated $\Delta$-sequence are given in Table \ref{tab:house_data}. Also, see Figure \ref{fig:house} to see typical fluctuations in the power consumption of the house and the generated $\Delta$-sequence.

Based on the data,  we choose $\epsilon$ to be 40 W ($\approx0.25\%$ of the maximum power usage). From our experiments (see Table \ref{tab:fp_house}), we observe that even for a very small sample size of 30,  the filtered $\Delta$ mean difference algorithm gives us a very good FP/FN rate of less than $1\%$,  i.e., with around 99.9 percent accuracy,  the algorithm can detect the presence of an attacker (if any) in less than 30 seconds.  %
This case also highlights the advantage of the filtration step that essentially removes the sudden changes in the power drawn when some appliance is turned on (or off).  We see that without such a filtration step, the FP/FN rate is around $4.2\%$ for a similar duration of 30 seconds. Notice from Table \ref{tab:house_data}, that the maximum $\Delta$ is quite large. Filtering these high $\Delta$ values reduces the variance of the considered $\Delta'$ sequence allowing for faster and more accurate detection. %
In contrast, the simple mean difference algorithm required around 20,000 samples to achieve a similar level of accuracy of less than $1\%$.

\begin{table}[]
\begin{center}
\begin{tabular}{|c|c|c|}
\hline
& Solar Power (kW) & $\Delta$-sequence (kW)\\ \hline
Maximum Value & 1576.54  &  1132.67\\ \hline
Minimum Value &  0.0  &  -925.18\\ \hline
Average Value & 394.118  &  $9.30 \times 10^{-5}$\\ \hline
Median Value &  276.36  &  -0.519\\ \hline
\end{tabular}
\end{center}
\caption{\small Table showing the power output data statistics for the solar plant and its associated $\Delta$-sequence.}
\label{tab:solar_data}
\end{table}

\subsection{Attack Detection in Solar Power Dataset}

\begin{figure}[h!]
  \centering
    \includegraphics[width=0.6\linewidth]{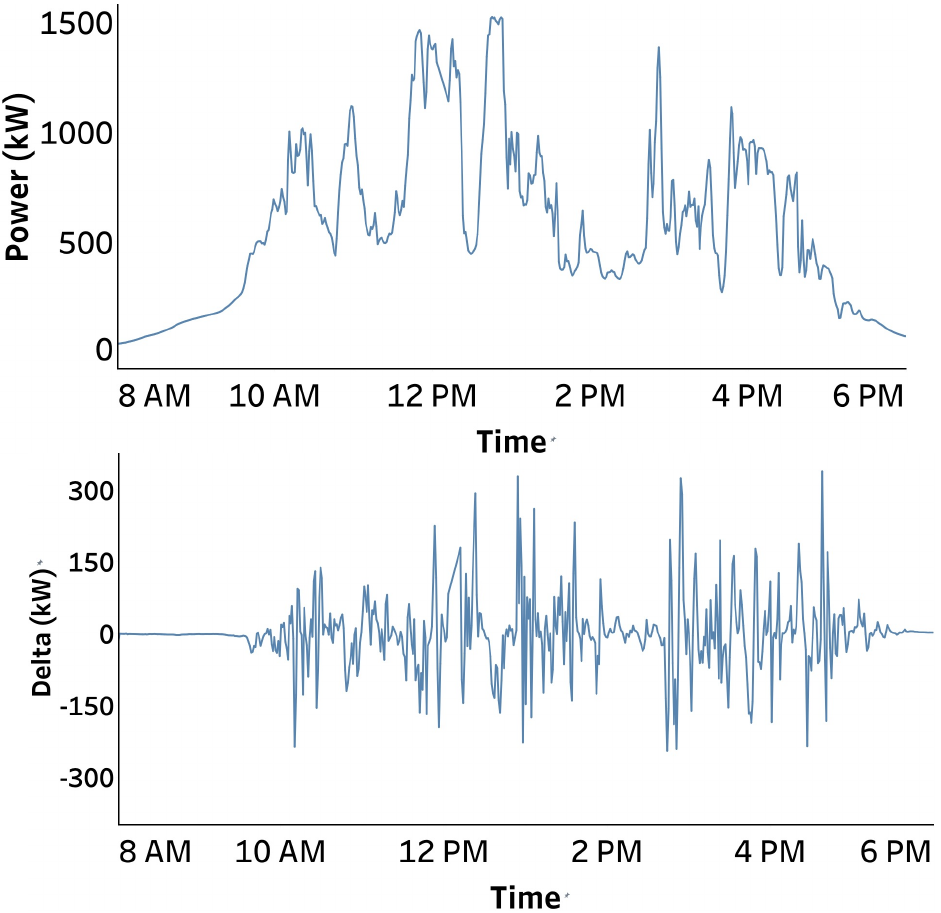}
  \caption{The solar power output and $\Delta_i$ fluctuation over a day.}
  \label{fig:solar}
\end{figure}

The solar power or the PV dataset is collected from a solar plant deployment in Singapore.  The data contains minute-wise values of the power generated from 7 stations from a period of 1/05/2020 to 17/06/2020, with the power generated given in kW for each station and the aggregate power output of the solar plant. 
We consider the sensor giving the aggregate power output of the solar plant to run our experiments.
The data statistics and the statistics of the associated $\Delta$-sequence are given in Table \ref{tab:solar_data}. Also, see Figure \ref{fig:solar} to see typical fluctuations in the solar data and the generated $\Delta$-sequence.

\begin{table}[h]

\centering
\resizebox{\columnwidth}{!}{
\begin{tabular}{|c|ccc|ccc|ccc|}
\hline
\multirow{3}{*}{n} & \multicolumn{3}{c|}{\begin{tabular}[c]{@{}c@{}}Simple Mean\\ Difference\end{tabular}}                                      & \multicolumn{3}{c|}{\begin{tabular}[c]{@{}c@{}}Mean\\ Difference\end{tabular}}                                          & \multicolumn{3}{c|}{\begin{tabular}[c]{@{}c@{}}Filtered\\ Mean Difference\end{tabular}}                                 \\ \cline{2-10} 
                   & \multicolumn{1}{c|}{\multirow{2}{*}{\begin{tabular}[c]{@{}c@{}}FP\\ (\%)\end{tabular}}} & \multicolumn{2}{c|}{FN (\%)}          & \multicolumn{1}{c|}{\multirow{2}{*}{\begin{tabular}[c]{@{}c@{}}FP\\ (\%)\end{tabular}}} & \multicolumn{2}{c|}{FN (\%)}       & \multicolumn{1}{c|}{\multirow{2}{*}{\begin{tabular}[c]{@{}c@{}}FP\\ (\%)\end{tabular}}} & \multicolumn{2}{c|}{FN (\%)}       \\ \cline{3-4} \cline{6-7} \cline{9-10} 
                   & \multicolumn{1}{c|}{}                                                                 & \multicolumn{1}{c|}{EDA}   & RDA   & \multicolumn{1}{c|}{}                                                                 & \multicolumn{1}{c|}{EDA} & RDA  & \multicolumn{1}{c|}{}                                                                 & \multicolumn{1}{c|}{EDA} & RDA  \\ \hline
30                 & \multicolumn{1}{c|}{69.82}                                                            & \multicolumn{1}{c|}{12.3}  & 12.99 & \multicolumn{1}{c|}{0.0}                                                              & \multicolumn{1}{c|}{0.0} & 1.79 & \multicolumn{1}{c|}{0.0}                                                              & \multicolumn{1}{c|}{0.0} & 1.29 \\ \hline
60                 & \multicolumn{1}{c|}{71.76}                                                            & \multicolumn{1}{c|}{13.39} & 13.59 & \multicolumn{1}{c|}{0.0}                                                              & \multicolumn{1}{c|}{0.0} & 0.20 & \multicolumn{1}{c|}{0.0}                                                              & \multicolumn{1}{c|}{0.0} & 0.12 \\ \hline
90                 & \multicolumn{1}{c|}{74.15}                                                            & \multicolumn{1}{c|}{13.78} & 13.84 & \multicolumn{1}{c|}{0.0}                                                              & \multicolumn{1}{c|}{0.0} & 0.01 & \multicolumn{1}{c|}{0.0}                                                              & \multicolumn{1}{c|}{0.0} & 0.01 \\ \hline
120                & \multicolumn{1}{c|}{75.83}                                                            & \multicolumn{1}{c|}{14.74} & 14.75 & \multicolumn{1}{c|}{0.0}                                                              & \multicolumn{1}{c|}{0.0} & 0.0  & \multicolumn{1}{c|}{0.0}                                                              & \multicolumn{1}{c|}{0.0} & 0.0  \\ \hline
600                & \multicolumn{1}{c|}{76.43}                                                            & \multicolumn{1}{c|}{18.36} & 18.47 & \multicolumn{1}{c|}{0.0}                                                              & \multicolumn{1}{c|}{0.0} & 0.0  & \multicolumn{1}{c|}{0.0}                                                              & \multicolumn{1}{c|}{0.0} & 0.0  \\ \hline
                   
\end{tabular}
}
\caption{\small Table showing the false positive and the false negative rate (expressed as percentage) for different $n$ for simple, $\Delta$ and filtered $\Delta$ mean difference algorithms done over 10,000 trials for the solar dataset with $\epsilon = 7.5$ kW.  For filtration $\Delta_{th} = 30$ kW.}
\label{tab:fp_solar}
\end{table}

 We know that at night time, the solar power output sensor readings are all zeros, where the presence of an attacker can be detected extremely fast. Hence, our evaluation only considers the more challenging daytime values (from $8$am to $6$pm), which is 10 hours, or 600 points (per-minute) with a chosen $\epsilon$ as $7.5$kW (or 0.5\% of the maximum output of around $1.5$MW in this solar plant). As such, for each algorithm, we evaluate at most 600 samples, after which we assume that the attacker can be detected. Even at daytime, we observe from Table \ref{tab:fp_solar} that we can detect an attacker with relatively high accuracy (of around 99.8$\%$) with 60 samples (that translates to 1hr time) for the $\Delta$ mean based algorithms both with and without the filtration step. In this case, as observed from Table \ref{tab:solar_data}, the maximum $\Delta$ is not that high, and therefore,  we see that though the filtration step helps, it does not have a significant impact on the accuracy as we saw earlier in Section \ref{sec:house}. In addition, we also see that the simple mean difference algorithm for these short intervals performs quite poorly.  %

%

%
%
%
%
%
%
%
%
%
%
%
%
%
%
%
%
%

%
%
%
%
%
%
%
%

\subsection{Attack Detection in SWaT Dataset} \label{sec:swat}
\begin{figure}[h!]
  \centering
    \includegraphics[width=0.6\linewidth]{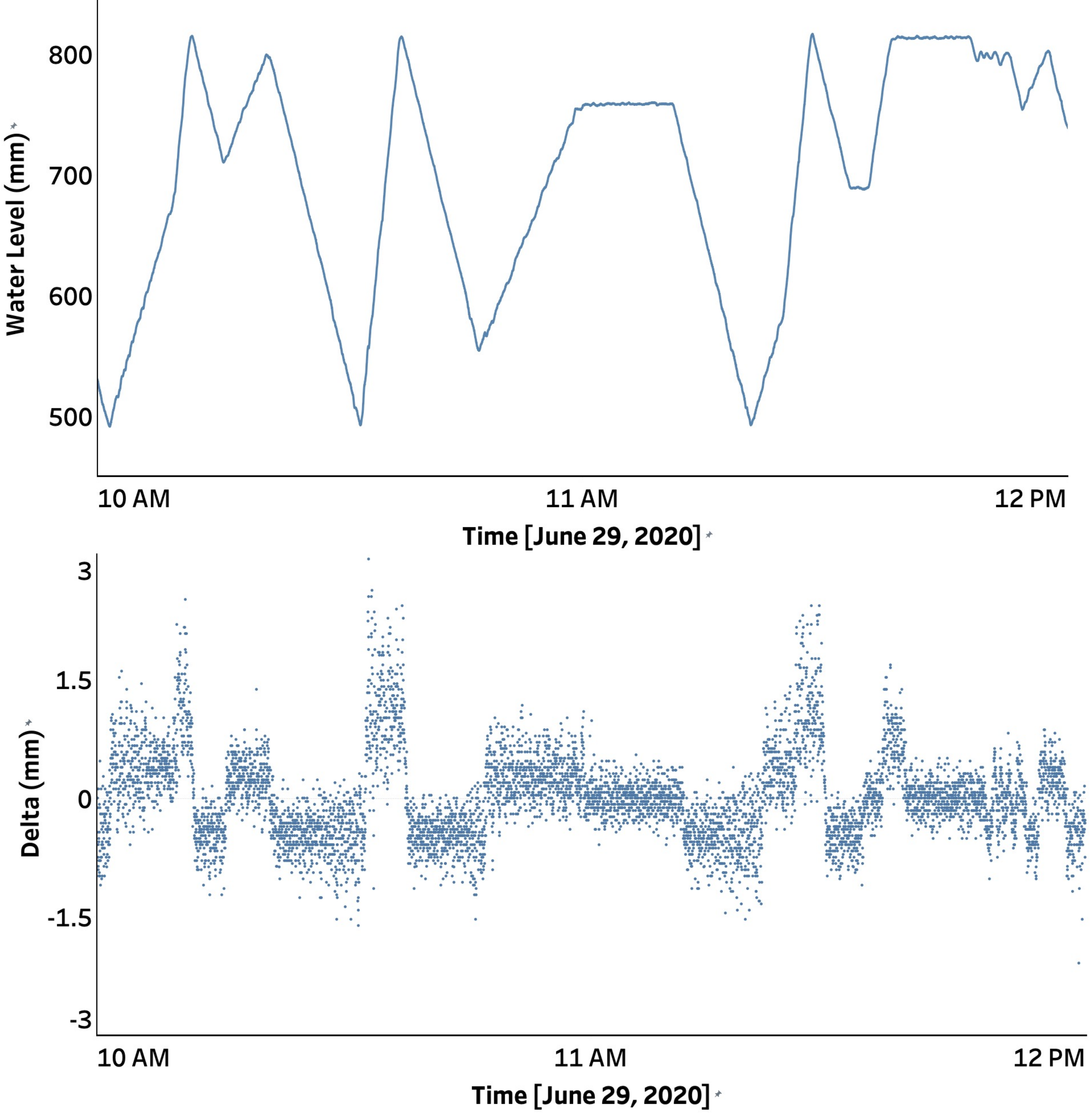}
  \caption{The water level and $\Delta_i$ fluctuation of raw water tank in the SWaT testbed.}
  \label{fig:swat}
\end{figure}

The Secure Water Treatment (SWaT) dataset is a publicly available dataset obtained from a state-of-the-art water treatment testbed %
\cite{10.1145/3274694.3274748}.  SWaT imitates the complete process of a real water treatment plant, and produces 5 gallons/minute filtered water. Overall, it contains second-wise data from 18 sensors  along with other status and state indicators. We consider the  measurements of sensor 'LIT101' (level indication transmitter) that records the second-wise water level of the raw water tank in millimeters on 29-05-2020, as the sensor under attack. %
The data statistics and the statistics of the associated $\Delta$-sequence are given in Table \ref{tab:swat_data}. Also, see Figure \ref{fig:swat} for typical fluctuations in the SWaT data and the generated 
$\Delta$-sequence.

\begin{table}[]
\begin{center}
\begin{tabular}{|c|c|c|}
\hline
& Water Level (mm) & $\Delta$-sequence (mm)\\ \hline
Maximum Value & 816.968  &  3.022\\ \hline
Minimum Value &  491.484  &  -2.080\\ \hline
Average Value & 698.777  &  0.029\\ \hline
Median Value &  727.118  &  0.0\\ \hline
\end{tabular}
\end{center}
\caption{\small Table showing the water level data statistics for the SWaT testbed and its associated $\Delta$-sequence.}
\label{tab:swat_data}
\end{table}

\begin{table}[h]
\centering
\resizebox{\columnwidth}{!}{
\begin{tabular}{|c|ccc|ccc|ccc|}
\hline
\multirow{3}{*}{n} & \multicolumn{3}{c|}{\begin{tabular}[c]{@{}c@{}}Simple Mean\\ Difference\end{tabular}}                                      & \multicolumn{3}{c|}{\begin{tabular}[c]{@{}c@{}}Mean\\ Difference\end{tabular}}                                           & \multicolumn{3}{c|}{\begin{tabular}[c]{@{}c@{}}Filtered\\ Mean Difference\end{tabular}}                                  \\ \cline{2-10} 
                   & \multicolumn{1}{c|}{\multirow{2}{*}{\begin{tabular}[c]{@{}c@{}}FP\\ (\%)\end{tabular}}} & \multicolumn{2}{c|}{FN (\%)}          & \multicolumn{1}{c|}{\multirow{2}{*}{\begin{tabular}[c]{@{}c@{}}FP\\ (\%)\end{tabular}}} & \multicolumn{2}{c|}{FN (\%)}        & \multicolumn{1}{c|}{\multirow{2}{*}{\begin{tabular}[c]{@{}c@{}}FP\\ (\%)\end{tabular}}} & \multicolumn{2}{c|}{FN (\%)}        \\ \cline{3-4} \cline{6-7} \cline{9-10} 
                   & \multicolumn{1}{c|}{}                                                                 & \multicolumn{1}{c|}{EDA}   & RDA   & \multicolumn{1}{c|}{}                                                                 & \multicolumn{1}{c|}{EDA}  & RDA  & \multicolumn{1}{c|}{}                                                                 & \multicolumn{1}{c|}{EDA}  & RDA  \\ \hline
30                 & \multicolumn{1}{c|}{2.78}                                                             & \multicolumn{1}{c|}{1.38}  & 3.08  & \multicolumn{1}{c|}{0.00}                                                             & \multicolumn{1}{c|}{0.00} & 3.37 & \multicolumn{1}{c|}{0.00}                                                             & \multicolumn{1}{c|}{0.00} & 3.42 \\ \hline
50                 & \multicolumn{1}{c|}{4.84}                                                             & \multicolumn{1}{c|}{2.39}  & 3.42  & \multicolumn{1}{c|}{0.00}                                                             & \multicolumn{1}{c|}{0.00} & 0.91 & \multicolumn{1}{c|}{0.00}                                                             & \multicolumn{1}{c|}{0.00} & 0.92 \\ \hline
60                 & \multicolumn{1}{c|}{6.25}                                                             & \multicolumn{1}{c|}{3.00}  & 3.85  & \multicolumn{1}{c|}{0.00}                                                             & \multicolumn{1}{c|}{0.00} & 0.59 & \multicolumn{1}{c|}{0.00}                                                             & \multicolumn{1}{c|}{0.00} & 0.57 \\ \hline
90                 & \multicolumn{1}{c|}{9.35}                                                             & \multicolumn{1}{c|}{4.63}  & 5.10  & \multicolumn{1}{c|}{0.00}                                                             & \multicolumn{1}{c|}{0.00} & 0.10 & \multicolumn{1}{c|}{0.00}                                                             & \multicolumn{1}{c|}{0.00} & 0.12 \\ \hline
120                & \multicolumn{1}{c|}{12.59}                                                            & \multicolumn{1}{c|}{6.00}  & 6.50  & \multicolumn{1}{c|}{0.00}                                                             & \multicolumn{1}{c|}{0.00} & 0.03 & \multicolumn{1}{c|}{0.00}                                                             & \multicolumn{1}{c|}{0.00} & 0.03 \\ \hline
150                & \multicolumn{1}{c|}{14.29}                                                            & \multicolumn{1}{c|}{6.93}  & 7.44  & \multicolumn{1}{c|}{0.00}                                                             & \multicolumn{1}{c|}{0.00} & 0.00 & \multicolumn{1}{c|}{0.00}                                                             & \multicolumn{1}{c|}{0.00} & 0.00 \\ \hline
600                & \multicolumn{1}{c|}{27.61}                                                            & \multicolumn{1}{c|}{12.57} & 12.50 & \multicolumn{1}{c|}{0.00}                                                             & \multicolumn{1}{c|}{0.00} & 0.00 & \multicolumn{1}{c|}{0.00}                                                             & \multicolumn{1}{c|}{0.00} & 0.00 \\ \hline
\end{tabular}
}
\caption{\small Table showing the false positive and the false negative rate (expressed as percentage) for different $n$ for simple, $\Delta$ and filtered $\Delta$ mean difference algorithms done over 10,000 trials for the SWaT dataset with $\epsilon = 4.08$ mm.  For filtration $\Delta_{th} = 25$ mm.}
\label{tab:fp_swat}
\end{table}

Based on the data,  we choose $\epsilon$ to be 4.08 mm ($\approx0.5\%$ of the maximum water level). From our experiments (see Table \ref{tab:fp_swat}), we observe that even for a very small sample size of 50,  the filtered $\Delta$ mean difference algorithm gives us a very good FP/FN rate of less than $1\%$,  i.e., with around 99.9 percent accuracy,  the algorithm can detect the presence of an attacker (if any) in less than 50 seconds.  %
  We see that the results with and without a filtration step, the FP/FN rate is almost identical for all cases, indicating that the filtration does not help in this case. This is because of the fact that most of the second-wise changes that are seen here are gradual (as reflected in the maximum and minimum $\Delta$ values in Table \ref{tab:swat_data}). %

\subsection{Attack Detection in Synthetic Dataset}

In this section, we analyze the performance of our algorithm in the worst and best-case scenarios. Of course, the best-case scenario is when the system is totally steady, i.e., all sensor readings are identical %
or uniformly changing by either increasing or decreasing at a constant rate. For the uniformly changing case, the $\Delta$ between consecutive readings becomes constant. In these cases, as discussed earlier, an attacker can be detected extremely fast.

In contrast, the worst-case scenario for our detection algorithm is the case where the sensor readings are uniformly distributed over a large range. The $\Delta'$ sequence generated for this case will also be uniformly distributed. %

We create a synthetic dataset where each element is drawn uniformly and independently at random from a range of (0-100), and the chosen value of $\epsilon$ is $0.5$. %
If we look at $1\%$ or $0.1\%$ as a target FP/FN rate, we observe that using the simple mean difference algorithm requires around $83,000$ and $150,000$ samples, respectively. That is about 1 day and almost 2 days, respectively, for a per-second sample. In comparison, achieving a similar result with the $\Delta$-mean difference algorithm requires around $195,000$ and $295,000$ samples respectively which translate to slightly more than 2 days and 3 days respectively for a per-second sample.  This confirms for uniformly random independently generated samples, using the original samples is better than using the $\Delta$-sequence. Intuitively, calculating the $\Delta$ results in a random variable within the range of $-100$ to $100$, while using the original is in the range of $0$ to $100$. Expansion of the range explains the larger variance, and in turn the larger detection time. See Table \ref{tab:fp_synthetic} for obtained FP/FN rate for different number of observations duration ($n$).

\begin{table}[]
\begin{center}
\resizebox{0.9\columnwidth}{!}{\begin{tabular}{|c|c|c|c|c|c|c|}
\hline
\multirow{2}{*}{$n$} & \multicolumn{3}{c|}{\begin{tabular}[c]{@{}c@{}}Simple \\ Mean Difference\end{tabular}} & \multicolumn{3}{c|}{\begin{tabular}[c]{@{}c@{}}$\Delta$\\ Mean Difference\end{tabular}} \\ \cline{2-7} 
& \begin{tabular}[c]{@{}c@{}}FP(\%)\end{tabular} & \begin{tabular}[c]{@{}c@{}}FN (\%)\\ (EDA)\end{tabular} & \begin{tabular}[c]{@{}c@{}}FN (\%)\\ (RDA)\end{tabular} & \begin{tabular}[c]{@{}c@{}}FP(\%)\end{tabular} & \begin{tabular}[c]{@{}c@{}}FN (\%)\\ (EDA)\end{tabular} & \begin{tabular}[c]{@{}c@{}}FN (\%)\\ (RDA)\end{tabular} \\ \hline
50000 & 5.87 & 2.11 & 2.02 & 17.22 & 8.51 & 8.54 \\ \hline
100000 & 0.85 & 0.63 & 0.51 & 5.5 & 2.32 & 2.40 \\ \hline
150000 & 0.09 & 0.04 & 0.10 & 1.8 & 1.07 & 1.06 \\ \hline
200000 & 0.01 & 0.00 & 0.01 & 0.90 & 0.45 & 0.45 \\ \hline
250000 & 0.00 & 0.00 & 0.00 & 0.22 & 0.20 & 0.23 \\ \hline
300000 & 0.00 & 0.00 & 0.00 & 0.00 & 0.00 & 0.00 \\ \hline
\end{tabular}}
\end{center}
\caption{\small Table showing the false positive and the false negative rate (expressed as percentage) for different $n$ for simple and $\Delta$ mean difference algorithms done over 10,000 trials for a worst-case synthetic dataset with $\epsilon = 0.5$.\vspace{5mm}}
\label{tab:fp_synthetic}
\end{table}

\section{Conclusion}
In this paper, we presented micro-distortion based detection algorithm that can help detect hidden attackers. It provides fast and accurate detection, and requires less changes to legacy systems. The key challenge is to integrate the distortions  based on the shared secret key into the sensor readings itself while keeping the distortions as small as possible in such a way that  it does not affect the overall system performance. 
Our approach ensures that none of the original receivers of the sensor reading needs to know the secret key of a sensor to make use of the sensor's reading. Not only zero upgrading or change is needed for those receivers, this also reduces the risk of leaking the secret key due to those legacy receivers being compromised. Our evaluation shows that physically-induced micro distortion works well for systems that are relatively steady (i.e., not highly volatile most of the time). However, for systems that undergo frequent rapid fluctuations, physically-induced distortion would require a longer duration to detect an attacker. In those settings, a digitally-induced distortion approach may be more suitable.

\section*{Acknowledgments}
This research is supported in part by the National Research Foundation, Singapore, under its National Satellite of Excellence Programme  “Design Science and Technology for Secure Critical Infrastructure” (Award Number: NSoE DeST-SCI2019-0008) and “Design Science and Technology for Secure Critical Infrastructure: Phase II” and in part by the SUTD Start-up Research Grant (SRG Award No: SRG ISTD 2020 157). Any opinions, findings and conclusions or recommendations expressed in this material are those of the author(s) and do not reflect the views of National Research Foundation, Singapore. 

\balance
\bibliographystyle{IEEEtran}
\begin{small}
\bibliography{bibliography}
\end{small}

\begin{table}[h!]

\begin{IEEEbiography}[{\includegraphics[width=1in,height=1.25in,clip,keepaspectratio]{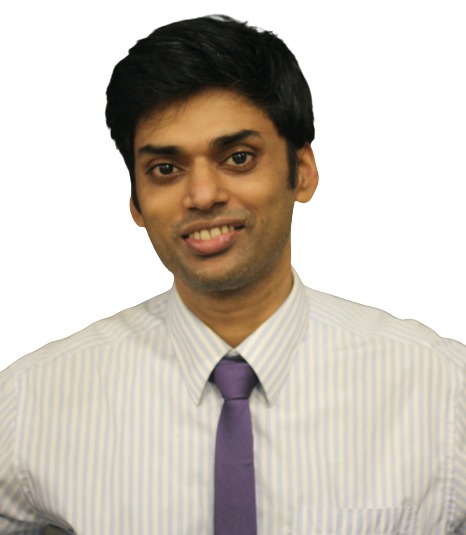}}]{Suman Sourav}
Suman Sourav is a postdoctoral researcher at the Singapore University of Technology and Design. He received his Ph.D. from the National University of Singapore where he also received the research achievement award for his work. He also worked as a research fellow at the Advanced Digital Sciences Centre which is the University of Illinois Urbana-Champaign's research division in Singapore. His research interest lies in cyber-physical system security, distributed computing and algorithms, both theoretical as well as practically implemented algorithmic aspects of real-world systems, which includes developing theoretical bounds and deployable protocols for both static and highly dynamic environments, achieving security in communication networks and cyber-physical systems, along with issues including robustness and scalability.
\end{IEEEbiography}

\begin{IEEEbiography}[{\includegraphics[width=1in,height=1.25in,clip,keepaspectratio]{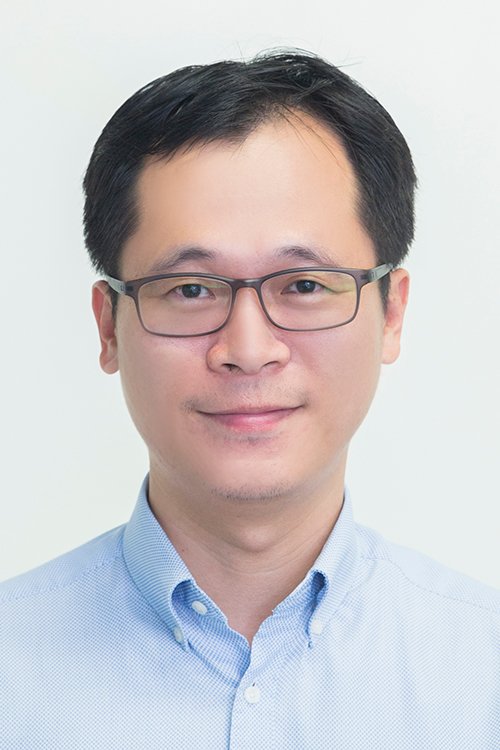}}]{Binbin Chen}
Binbin Chen (M'11) received the B.Sc. degree in computer science from Peking University and the Ph.D. degree in computer science from the National University of Singapore. Since July 2019, he has been an Associate Professor in the Information Systems Technology and Design (ISTD) pillar, Singapore University of Technology and Design (SUTD). His current research interests include wireless networks, cyber-physical systems, and cyber security for critical infrastructures.
\end{IEEEbiography}

\end{table}

\end{document}